\documentclass[aip,superscriptaddress,reprint, twocolumn, amsmath, amssymb, a4, notitlepage]{revtex4-2}
\usepackage{graphicx}
\usepackage{amsmath,amssymb}
\usepackage{dcolumn}
\usepackage{mathrsfs}
\usepackage{physics}
\usepackage[utf8]{inputenc}
\usepackage{float}
\usepackage{booktabs}
\usepackage{svg}
\usepackage{comment}
\usepackage{color}
\usepackage{hyperref}

\newcommand{\trS}[1]{\mathrm{tr}_{\mathrm{S}}\left[#1\right]}
\newcommand{\trB}[1]{\mathrm{tr}_{\mathrm{B}}\left[#1\right]}
\newcommand{\trSB}[1]{\mathrm{tr}_{\mathrm{S+B}}\left[#1\right]}

\begin{document}
\title{Beyond the Quantum Regression Theorem in Variational Polaron Master Equations: from Ohmic to Super-Ohmic Environments}
\date{\today}
	
\author{Matias Bundgaard-Nielsen}
\affiliation{Department of Electrical and Photonics Engineering, Technical University of Denmark, Building 343, 2800 Kongens Lyngby, Denmark}
\affiliation{NanoPhoton-Center for Nanophotonics, Technical University of Denmark, Building 343, 2800 Kongens Lyngby, Denmark}
\author{Jake Iles-Smith}
\affiliation{School of Mathematical and Physical Sciences, The University of Sheffield, Western Bank, Sheffield, S10 2TN, United Kingdom}
\email{jake.iles-smith@sheffield.ac.uk}

\begin{abstract}
While the quantum regression theorem (QRT) is the standard tool for computing multi-time correlation functions in open quantum systems, it relies on system–bath separability and an environment that remains in equilibrium, assumptions that are violated once dynamical correlations develop. Using the projection operator formalism, we derive an extension to the QRT that explicitly incorporates these correlation-induced corrections, enabling us to capture multi-time correlation functions in non-Markovian regimes.
We apply this framework to the variational polaron master equation for the spin–boson model in ohmic and super-ohmic regimes, where the polaron transformation mixes system–bath degrees of freedom to produce a non-thermal effective environment. Benchmarking against numerically exact tensor-network simulations, we demonstrate quantitative agreement for single- and two-time observables, including linear-response spectra, even at strong coupling. 
Our approach broadens the reach of analytic master equations to multi-time observables in strong-coupling and non-Markovian regimes, where environmental memory effects and system–bath correlations are crucial.

\end{abstract}
\maketitle

\section{Introduction}
System–environment interactions play a decisive role in the dynamics of molecular and solid-state systems, governing fundamental processes such as vibrational relaxation, electronic dephasing, and charge transport. Despite their importance, accurately modeling these open quantum systems remains a significant challenge due to the complex, non-Markovian nature of condensed-phase environments. Recently, numerically exact methods based on tensor networks—for example, those utilizing the process tensor formalism \cite{Pollock2018OperationalProcesses, Strathearn2018EfficientOperators, Jrgensen2019ExploitingIntegrals, Cygorek2022SimulationEnvironments, Cygorek2024ACE} or chain mapping approaches \cite{Prior2010,Schrder2016,Schrder2019,Tamascelli2019}—have emerged as transformative tools for simulating these dynamics. These approaches have enabled highly accurate simulations in previously inaccessible regimes, particularly those characterized by strong system–bath coupling and structured spectral densities \cite{Svendsen2023SignaturesMaterials, FerreiraNeto2024One-dimensionalDimensionality}. However, while tensor-network methods are increasingly efficient, they often function as a black box that can be challenging to interpret physically. Furthermore, their unfavorable computational scaling with system size and complexity highlights the continued need for transparent, analytically grounded frameworks that provide physical insight alongside numerical reliability.

In contrast, perturbative master equations offer a conceptually transparent alternative by deriving reduced dynamics through systematic expansions \cite{Breuer2006TheSystems}. While traditionally limited by weak-coupling or Markovian assumptions \cite{Wrger1998,McCutcheon2010QuantumRegime, McCutcheon2011ADots,Kaer2013MicroscopicDecoherence,Bundgaard-Nielsen2021Non-MarkovianElectrodynamics}, these methods provide a clear mapping between mathematical terms and physical processes that is often obscured in purely numerical treatments. By employing unitary mappings to incorporate environmental degrees of freedom into an augmented system, as in polaron and reaction-coordinate master equations \cite{Wilson-Rae2002QuantumInteractions,Hughes2011, McCutcheon2011ADots,Roy2011Phonon-dressedEffects, Hornecker2017InfluenceFunctions, Denning2020OpticalInteractions, Bundgaard-Nielsen2021Non-MarkovianElectrodynamics} or through alternative partitions in which off-diagonal electronic couplings are treated perturbatively, as in Förster and modified-Redfield master equations \cite{Trushechkin2019CalculationTransfer,Lai2021OnPerturbation}, perturbative frameworks can be extended into strong-coupling regimes. This enables a detailed understanding of the dynamics, allowing specific corrections, such as the non-equilibrium environmental effects addressed here, to be formally isolated and analyzed.

\begin{figure*}
    \centering
    \includegraphics[width=\linewidth]{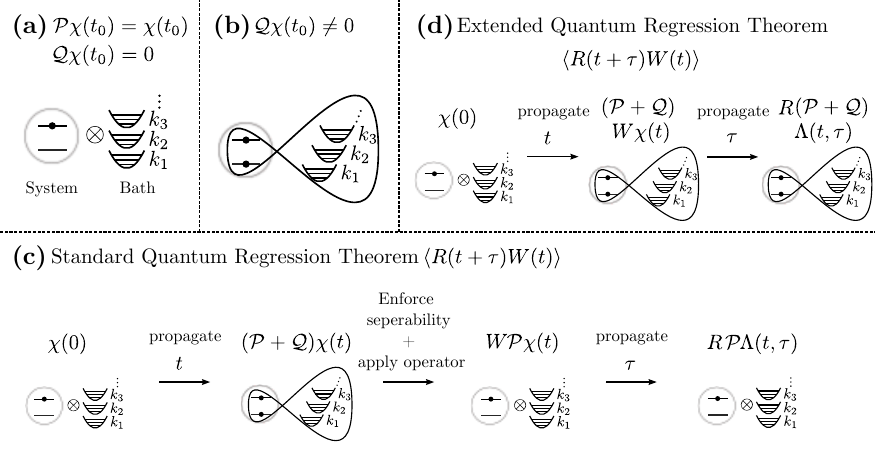}
    \caption{Schematic illustration of the standard and extended quantum regression constructions.
    (a) An initially factorized system--bath state, which is the reference state generated by $\mathcal{P} \chi(t_0)$ and has $\mathcal{Q}\chi(t_0)=0 $.
    (b) A system and bath that is initially entangled / out of equilibrium and thus $\mathcal{Q} \chi(t_0)\neq0$
    (c) In the standard quantum regression theorem (QRT), the system is first propagated to time $t$, an operator $W$ is applied, and the resulting auxiliary state is projected onto the reference state by $\mathcal{P}$ and then propagated for a further delay time $\tau$ before measuring an operator $R$. This construction effectively keeps only the factorized contribution $\mathcal{P}$ and neglects the $\mathcal Q$ component generated during the first propagation.
    (d) In the extended QRT used here, the auxiliary state after the action of $W$ contains both $\mathcal P \chi$ and $\mathcal Q\chi$ components. Propagating both contributions from $t$ to $t+\tau$ accounts for system--bath correlations and non-equilibrium bath contributions that are absent in the standard QRT.    \label{fig:QRT}}
\end{figure*}

Furthermore, polaron master equations have been widely used to include non-Markovian bath effects in a Markovian master equation in a transformed frame \cite{Wilson-Rae2002QuantumInteractions,Hughes2011, McCutcheon2011ADots,Roy2011Phonon-dressedEffects, Hornecker2017InfluenceFunctions, Denning2020OpticalInteractions, Bundgaard-Nielsen2021Non-MarkovianElectrodynamics}. This is particularly useful for multi-time correlations such as linear response functions and spectra calculated using the quantum regression theorem (QRT). When system operators are transformed back to the laboratory frame, bath displacement operators appear, and their correlation functions generate phonon sidebands and other memory-induced spectral features \cite{Iles-Smith2017PhononSources}. In this way, polaron master equations used together with the QRT can calculate multi-time correlation functions containing non-Markovian environmental dressings. 

However, the standard polaron transformation is not well behaved for spectral densities with strong low-frequency weight. For spectral densities of the form $J(\omega)\propto\omega^s$, it suffers from infrared divergences for $s\leq 2$ \cite{Silbey1983VariationalBath, Chassagneux2018EffectRegime, FerreiraNeto2024One-dimensionalDimensionality}, where for deformation-potential coupling the exponent $s$ can be related to the dimensionality of the phononic environment \cite{Chassagneux2018EffectRegime} and a large class of environments are therefore not treatable using polaron approaches.

The variational polaron transformation alleviates this problem by optimizing the displacement of each bath mode \cite{Silbey1983VariationalBath, Harris1985VariationalSystem,Silbey1989TunnelingDescription}. In this way, it interpolates between the weak-coupling and fully displaced polaron limits \cite{McCutcheon2011ADots,Hsieh2019ATransport,Wang2020VariationalCrystals,Denning2020OpticalInteractions,Bundgaard-Nielsen2021Non-MarkovianElectrodynamics}. Slow modes that cannot adiabatically follow the system are displaced less strongly, while modes for which polaron dressing is favorable remain included in the transformed system description. This makes the variational approach a natural candidate for treating low-dimensional phonon environments. Nevertheless, its accuracy in regimes with strong coupling to slow environmental modes remains a relatively unexplored area, especially for multi-time correlation functions.

A further subtlety arises because the polaron transformation mixes system and environmental degrees of freedom. Recent work has shown that standard polaron master equations can accurately describe observables that commute with the transformation, such as populations in the spin--boson model, while coherences require additional correction terms associated with the $\mathcal Q$-projected component of the transformed density matrix \cite{Iles-Smith2024CapturingFormalism}. These terms describe the dynamical dressing of the system by the environment and can affect both transient and steady-state coherences. This also highlights the importance of the environmental state itself. In variational polaron treatments, the initial state of the bath is typically taken as the thermal equilibrium state of the displaced bath; however, because it is defined in
the transformed frame, it does not correspond to a simple Gibbs state in the laboratory frame and thus implicitly encodes non-trivial system–environment correlations already at the initial time. This gives rise to inhomogeneous terms in the master equation and to correction terms that can contribute directly to observable quantities \cite{McCutcheon2011ConsistentEquation}.

In this work, we derive and benchmark a variational polaron master equation for low-dimensional phonon environments, with particular focus on these correlation-induced corrections. Using the projection-operator formalism \cite{nakajima1958quantum,zwanzig1960ensemble,Shibata1977,Breuer2006TheSystems,mccutcheon_optical16,Iles-Smith2024CapturingFormalism}, we obtain a time-convolutionless master equation together with correction terms to expectation values with operators that do not commute with the variational polaron transformation. For one-time observables, these terms correct laboratory-frame expectation values. For two-time correlation functions, they lead to an extended QRT, which includes additional inhomogeneous source terms as well as corrections to the expectation value. This is also illustrated in Fig.\ref{fig:QRT}.

We apply this framework to the spin–boson model, a ubiquitous paradigm in chemical and condensed matter physics. Both the standard and variational polaron master equations have been used extensively to model this system across a broad range of phenomena, spanning from excitation energy transfer in photosynthetic complexes \cite{Akihito2009AdequacyTransfer, Leonardo2011PhysicalSystems} and electron transport in organic crystals \cite{Wang2020VariationalCrystals}, to nanoscale heat transport \cite{Hsieh2019ATransport}, condensed-phase tunneling \cite{Leggett1987Dynamics}, and decoherence in solid-state emitters \cite{Nazir2016ModellingDots, Iles-Smith2017PhononSources, Iles-Smith2017LimitsEmitters, Denning2020PhononSources, Bundgaard-Nielsen2021Non-MarkovianElectrodynamics}, as well as broader quantum phase transitions \cite{Chin2006CoherentModel, Chin2011GeneralizedTransition, Tong2011QuantumModel}. We systematically investigate spectral densities characterized by different low-frequency power-law scalings, $J(\omega) \propto \omega^s$, paying particular attention to the cases where the dimensionality $s\leq2$ and the standard polaron transformation notoriously suffers from infrared divergences \cite{Silbey1983VariationalBath, Chassagneux2018EffectRegime, FerreiraNeto2024One-dimensionalDimensionality}. With the recent interest in quantum emitters formed from defects in 2D materials \cite{Wang2016CoherentTemperature, Han2018RabiTemperature,Selig2016ExcitonicDichalcogenides,Fischer2023CombiningOrigin}, quantum dots embedded in 1D nanophotonic wires \cite{FerreiraNeto2024One-dimensionalDimensionality}, or trapped excitons in single carbon nanotubes (1D) \cite{Watahiki2012Enhancementnanocavities,Jeantet2016WidelyRegime,Jeantet2017ExploitingNanotube}, understanding how dimensionality impacts the system dynamics is increasingly relevant. In this context, assessing the accuracy of variational polaron master equations across these regimes is essential.

We benchmark the theory against numerically exact tensor-network simulations. This allows us to assess both the accuracy of the variational polaron transformation itself and the importance of the correction terms. We first examine population dynamics and coherences, where the corrections reveal how system--bath correlations contribute to steady-state expectation values. We then turn to the linear-response spectrum, which provides a sensitive test of the non-Markovian two-time dynamics. In this case, the conventional variational-polaron QRT already contains phonon memory through the transformed operators, but we find that the additional non-equilibrium bath corrections are required to recover the correct response in the regimes considered. We further note that the linear response spectrum is an experimentally accessible way to directly probe the non-Markovianity of the system considered, as a purely Markovian model predicts a vanishing spectrum. 

Our results show that the variational polaron transformation gives an accurate description of the population dynamics over a broad range of parameters, including regimes where the standard polaron transformation is ill-defined. For super-ohmic environments, the corrected theory reproduces coherences and linear-response spectra with excellent accuracy, where the importance of the corrections increases for smaller dimensions (decreasing $s$). For strictly ohmic environments ($s=1$), however, strong coupling to slow modes limits the accuracy of the predicted coherences, even when the populations remain well described. At the same time, comparison with tensor-network simulations provides insight into the physical origin of the observed dynamics and helps make aspects of the black box numerical calculations transparent, such as the need to "relax" the tensor-network state in order to simulate dynamics with an equilibrium bath state.

The remainder of this paper is organized as follows. Sec.~\ref{sec:me} details the projection operator derivation of the second-order master equation and the extended quantum regression theorem. Sec.~\ref{sec:model_and_var} introduces the spin–boson model and the variational polaron transformation. In Sec.~\ref{sec:corrections}, we present the specific expectation value corrections arising from system–bath mixing. Sections \ref{sec:populations} and \ref{sec:linear_response} provide benchmarking against exact tensor-network simulations for system dynamics and linear-response spectra, respectively. Finally, we conclude and provide an outlook in Sec.~\ref{sec:conclusion}.

\section{Master equation \label{sec:me}}
In this section, we present the derivation of the second‑order Time Convolutionless (TCL) master equation (ME), including the often-omitted inhomogeneous term that arises when the initial bath is not in a thermal equilibrium state. As we will see, the inclusion of inhomogeneous terms also has a significant impact on the quantum regression theorem \cite{Breuer2006TheSystems}, where we find corrections even if the initial state is thermal.

\subsection{The Projection operator formalism}
We wish to describe the dynamics of the reduced system state in the interaction picture (denoted by a tilde) $\tilde{\rho}_S(t)=\mathrm{tr}_B(\tilde{\chi}(t))$, which is initially in a product state with the environment: ${\chi}(t_0)={\rho}(t_0)\otimes \tau_B$, where $\rho(t_0)$ is the initial state of the system and $\tau_B$ the initial state of the bath. To do so, we introduce the projection operators~\cite{nakajima1958quantum, zwanzig1960ensemble},
\begin{equation}\label{eq:projectors}
    \mathcal{P}{\chi} \equiv \trB{\chi}\otimes \tau_R,\qquad \mathcal{Q}=1-\mathcal{P}.
\end{equation}
Here $\tau_R$ is a reference state of the environment and is often taken to be the Gibbs state $\tau_{\mathrm{R}}=\exp \left(-\beta H_B\right) / \operatorname{tr}_{\mathrm{B}}\left(\exp \left(-\beta H_B\right)\right)$ with $H_B$ being the bare Hamiltonian of the bath, $\beta = 1/(k_B T)$, $T$ being the temperature, and $k_B$ Boltzmann’s constant. Typically, one also chooses the initial state such that $\tau_B = \tau_R$, meaning that $\mathcal{Q} {\chi}(t_0) = 0$. However, as we will see in the following section, when one has moved into the variational polaron frame, this will generally not be the case.

Following typical TCL ME derivations, see e.g., refs.~\cite{McCutcheon2011ConsistentEquation,Iles-Smith2024CapturingFormalism}, we consider the evolution of the total density operator in an interaction picture with respect to $H_0 = H_S + H_B$ ($H_S$ being the bare Hamiltonian of the system):
\begin{equation}\label{eq:vonN}
\partial_{t} \tilde{\chi}(t)=-i\left[{H}_{\mathrm{I}}(t), \tilde{\chi}(t)\right] \equiv \lambda \mathcal{L}(t) [\tilde{\chi}(t)] .
\end{equation}
where $\mathcal{L}(t)[\cdot]$ is the superoperator $\mathcal{L}(t)[ \tilde{\chi}(t)] = -i\left[{H}_{\mathrm{I}}(t), \tilde{\chi}(t)\right]$, $H_I$ the interaction Hamiltonian between the system and bath, and $\lambda$ represents a parameter about which we do perturbation. In the following, we denote all superoperators with calligraphic letters followed by square brackets surrounding the object they act on ($\mathcal{L}(t)[ \tilde{\chi}(t)]$, for example).  Applying the projection operators to Eq.~\ref{eq:vonN}, we obtain the equations of motion~\cite{Breuer2006TheSystems,Iles-Smith2024CapturingFormalism}:
\begin{align}
& \partial_t \mathcal{P} \tilde{\chi}(t)=\lambda \mathcal{P} \mathcal{L}(t) [\mathcal{Q} \tilde{\chi}(t)], \label{eq:P_EOM} \\
& \partial_t \mathcal{Q} \tilde{\chi}(t)=\lambda \mathcal{Q} \mathcal{L}(t) [\mathcal{P} \tilde{\chi}(t)]+\lambda \mathcal{Q} \mathcal{L}(t) [\mathcal{Q} \tilde{\chi}(t)], \label{eq:Q_EOM}
\end{align}
where, without loss of generality, we have assumed $\mathrm{tr}_B(H_I \tau_R)=0$.
We can formally integrate Eq.~\ref{eq:Q_EOM}, before expanding up to first order in $\lambda$, to obtain \cite{McCutcheon2011ConsistentEquation,Iles-Smith2024CapturingFormalism}:
\begin{equation}
\begin{aligned}
\mathcal{Q} \tilde{\chi}(t)&=  \mathcal{Q} {\chi}\left(t_0\right)+\lambda \int_{t_0}^t \mathrm{~d} s \mathcal{Q} \mathcal{L} \left(s\right) [\mathcal{Q} {\chi}\left(t_0\right)] \\
&+\lambda \int_{t_0}^t \mathrm{~d} s \mathcal{L}(s) [\mathcal{P} \tilde{\chi}(t)]+\mathcal{O}\left(\lambda^2\right) \label{eq:q_eom}
\end{aligned}
\end{equation}
Substituting Eq.~\eqref{eq:q_eom} into ~\eqref{eq:P_EOM}, yields an equation of motion for $\mathcal{P} \tilde{\chi}(t)$ valid to second order in $\lambda$,
\begin{equation}
\begin{aligned}
    \partial_t  \mathcal{P} \tilde{\chi}(t) =  \mathcal{K}(t,t_0)[ \mathcal{P} \tilde{\chi}(t)] + \mathcal{I}(t,t_0)[\mathcal{Q} {\chi}(t_0)]. \label{eq:me_total}   
\end{aligned}
\end{equation}
where the reduced density operator is obtained by tracing over the environment:
\begin{equation}
\begin{aligned}
    \partial_t \tilde{\rho}(t) =  \trB{ \mathcal{K}(t,t_0)[ \mathcal{P} \tilde{\chi}(t)]} + \trB{\mathcal{I}(t,t_0)[\mathcal{Q} {\chi}(t_0)]}. \label{eq:me_total_reduced}    
\end{aligned}
\end{equation}
This equation has two contributions. The first, termed the memory kernel, takes the form, 
\begin{equation}
    \mathcal{K}(t,t_0)[ \mathcal{P} \tilde{\chi}(t)] = - \lambda^2\int_{t_0}^{t}\!ds\; \bigl[{H}_\mathrm{I}(t),[{H}_\mathrm{I}(s), \mathcal{P} \tilde{\chi}(t)]\bigr], \label{eq:phonon_prop}
\end{equation}
The second term is referred to as the inhomogeneous term and arises only when $\mathcal{Q} {\chi}(t_0) \neq 0$,
and takes the form
\begin{align}
    \mathcal{I}(t,t_0)[\mathcal{Q} {\chi}(t_0)] &= -i\lambda [{H}_\mathrm{I}(t),\mathcal{Q}{\chi}(t_0)] \nonumber\\
    &-  \lambda^2\int_{t_0}^{t}\!ds\;\bigl[{H}_\mathrm{I}(t),[{H}_\mathrm{I}(s),\mathcal{Q}{\chi}(t_0)]\bigr]. \label{eq:general_inhomo_me}
\end{align}
If we assume that the interaction Hamiltonian can be decomposed into Hermitian operators,
$
H_I = \sum_i A_i \otimes B_i \label{eq:Hamiltonian_split}
$
where $A_i$ is a system operator and $B_i$ is a bath operator then we get with $\chi(t_0) = \rho_\mathrm{S}(t_0) \otimes \tau_B$ (taking $\lambda=1$):
\begin{align}
    \trB{ \mathcal{K}(t,t_0)[ \mathcal{P} \tilde{\chi}(t)]} = &- \sum_{ij}\int_{t_0}^t   \mathrm{d}s  \Big  \{ C_{ij}(s)[A_i,{A}_j(-s)\tilde{\rho}_{\rm S}(t)] \\
    &+ C_{ji}(-s)[\tilde{\rho}_{\rm S}(t) {A}_j(-s),A_i ]\Big\},\nonumber
\end{align}
where we defined $C_{ij}(\tau) =  \mathrm{tr}_B \left [ B_i(\tau)B_j(0) \tau_R \right ]$ and similarly:
\begin{equation}
\begin{aligned}
    &\trB{\mathcal{I}(t,t_0)[\mathcal{Q} {\chi}(t_0)]} = -i \sum_i  \Gamma_i(t) [A_i(t),\rho_\mathrm{S}(t_0)]  \\ 
    &- \sum_{ij}\int_{t_0}^t  \mathrm{d}   s\Big\{C^{\mathcal{I}}_{ij}(t,s) [A_i(t),{A}_j(s)  \rho_\mathrm{S}(t_0)] +\mathrm{h.c.} 
     \Big\}, \label{eq:me_inhomo}
\end{aligned}
\end{equation}
where we here introduced $\Gamma_i(t) = \tr_\mathrm{B} [ B_i(t) (\tau_B - \tau_R)]$ and $C_{ij}^\mathcal{I}(t,s) = \tr_\mathrm{B} [B_i(t)B_j(s) (\tau_B - \tau_R)]$.

\subsection{Implications for the Quantum Regression Theorem \label{sec:me_for_two_time}}
The master equation in Eq.~\ref{eq:me_total} is time-local in the reduced state, but its generator and inhomogeneous source term contain integrals over bath correlation functions. These time-dependent coefficients therefore encode memory of the bath dynamics, even though the equation itself is written in TCL form. Consequently, two-time correlation functions, such as $\expval{R(t+\tau)W(t)}$, cannot in general be obtained by applying the conventional quantum regression theorem (QRT) \cite{Breuer2006TheSystems}, which assumes that the auxiliary state after the operator insertion can be propagated with the same reduced equation of motion and without retaining the correlated $\mathcal Q$ component~\cite{mccutcheon_optical16,Cosacchi2021AccuracyDot,salamon2025markovian}. To extend the QRT, we must include the effects of the inhomogeneous evolution and the accumulation of system-environment correlations, as shown schematically in Fig.~\ref{fig:QRT}. 
We denote by $\mathcal{V}(t,t_0)$ the approximate evolution map that propagates an initial total system--bath operator $\chi(t_0)$ to $\chi(t)$ according to the projection-operator expansion above,
\begin{equation}
    \chi(t) = \mathcal{V}(t,t_0)[\chi(t_0)].
\end{equation}
More explicitly, the projected component $\mathcal{P}\chi(t)$ is obtained from Eq.~\eqref{eq:me_total}, while the irrelevant component $\mathcal{Q}\chi(t)$ is reconstructed perturbatively using Eq.~\eqref{eq:q_eom}. Since the evolution contains the inhomogeneous source term $\mathcal{I}(t)[\mathcal{Q}\chi(t_0)]$, the map depends on the full initial operator $\chi(t_0)$, not only on its projected part.

Using this propagator, we define a modified QRT:
\begin{equation}
\begin{aligned}
    \expval{R(t+\tau) W(t)} = \trSB{R(\tau) \tilde{\Lambda}(t,\tau)}
\end{aligned}
\end{equation}
where we defined:
\begin{equation}
    \Lambda(t,\tau) = \mathcal{V} (\tau,0) \left [ W \mathcal{V} (t,0) [\chi(0)] \right].
\end{equation}
Here, the second argument of $\Lambda(t,\tau)$ denotes the incremental time after the action of $W$ at time $t$. In other words, for the propagation of $\Lambda(t,\tau)$, we reset the time origin such that the initial time is implicitly $t_0=0$. The interaction picture is therefore defined with respect to this incremental time,
\begin{equation}
    \tilde{\Lambda}(t,\tau) = \mathrm{e}^{i H_0 \tau} \Lambda(t,\tau) \mathrm{e}^{- i H_0 \tau}.
\end{equation}
The initial state is here,
\begin{equation}
    \Lambda(t,0) = W \mathcal{V}(t,0)[ \chi(0)]= W (\mathcal{P} + \mathcal{Q}) \chi(t), \label{eq:xi_t}
\end{equation}
which we propagate for an incremental time $\tau$ and take the expectation value with $R$. The evolution is thus governed by $\mathcal{V}(\tau,0)[\Lambda(t,0)] = \mathcal{V}(\tau,0)[W (\mathcal{P} + \mathcal{Q})\chi(t)]$. Typically, in the QRT, the contribution of $W \mathcal{Q} \chi(t)$ is neglected, which can cause it to produce incorrect and unphysical results~
\cite{mccutcheon_optical16,Cosacchi2021AccuracyDot}. 
Furthermore, when $W$ acts on both the system and bath degrees of freedom (as it would in the polaron transformation~\cite{Iles-Smith2017LimitsEmitters,Iles-Smith2017PhononSources}), 
inhomogeneous corrections will also occur from $\mathcal{Q} W \mathcal{P} \chi(t)$, which are also typically neglected. 
As we will see in Sec.~\ref{sec:linear_correction}, these contributions play a significant role when, for example, calculating the linear response spectrum using quantum master equations.  

By applying the identity $\mathcal{P} + \mathcal{Q}$ to the initial state $\Lambda(t,0)$, using Eq.~\eqref{eq:me_total}, we get:
\begin{equation}
\begin{aligned}
    \partial_\tau \mathcal{P} \tilde{\Lambda}(t,\tau) 
    &= \mathcal{K}(\tau,0)[\mathcal{P}\tilde{\Lambda}(t,\tau)] 
    + \mathcal{I}(\tau,0)[\mathcal{Q}\Lambda(t,0)]. \label{eq:QRT}
\end{aligned}    
\end{equation}
Similarly, we can use Eq.~\eqref{eq:q_eom} to find:
\begin{equation}
\begin{aligned}
    \mathcal{Q} \tilde{\Lambda}(t,\tau) 
    &= \mathcal{Q} {\Lambda}(t,0) 
    + \lambda \int_{0}^{\tau} \mathrm{d} s \,
    \mathcal{Q} \mathcal{L}\left(s\right) [\mathcal{Q} {\Lambda}(t,0)] \\
&+\lambda \int_{0}^{\tau} \mathrm{d} s \,
    \mathcal{L}(s) [\mathcal{P} \tilde{\Lambda}(t,\tau)]
    +\mathcal{O}\left(\lambda^2\right). \label{eq:QLambda}
\end{aligned}
\end{equation}
With this we can thus find $\expval{R(t+\tau)W(t)} = \trSB{R (\mathcal{P}+\mathcal{Q})\tilde{\Lambda}(t,\tau)}$. In Sec.~\ref{sec:linear_correction}, we explicitly calculate the two-time correlation function for a linear-response spectrum.

\section{Variational polaron transformation \& the spin boson model \label{sec:model_and_var}}
We consider the spin–boson Hamiltonian, which describes a two-level system (TLS), with the two states $\ket{0}$ and $\ket{1}$, coupled to a bosonic environment. This model is widely used to capture the interaction of localized electronic degrees of freedom with vibrational environments, including, for example, phonon coupling in semiconductor quantum dots \cite{Nazir2016ModellingDots,Iles-Smith2017PhononSources,Iles-Smith2017LimitsEmitters,Denning2020PhononSources,Bundgaard-Nielsen2021Non-MarkovianElectrodynamics} as well as vibrational environments in molecular and chemical systems. The Hamiltonian for the unbiased spin-boson model reads \cite{Mahan2000Many-ParticlePhysics,Breuer2006TheSystems}:
\begin{equation}
    H = 
     \frac{\Delta}{2}\sigma_x + \sigma_z \sum_\mathbf{k} g_\mathbf{k}(b_\mathbf{k}^\dagger + b_\mathbf{k}) + \sum_\mathbf{k} \nu_\mathbf{k} b_\mathbf{k}^\dagger b_\mathbf{k} \label{eq:H}
\end{equation}
where we have introduced the Pauli operators $\sigma_z = \dyad{0}  - \dyad{1}$ and $\sigma_x = \dyad{1}{0} + \dyad{0}{1}$. 
Here, $\Delta$ denotes the tunneling rate or for a quantum emitter the strength of the laser coherently driving its transitions, $b_\mathbf{k}$ ($b_\mathbf{k}^\dagger$) the annihilation (creation) operator of the bosonic mode with momentum $\mathbf{k}$, frequency $\nu_\mathbf{k}$, and coupling strength $g_\mathbf{k}$. 
The overall strength of the interaction with the bosonic environment is characterized by the spectral density $J(\nu) = \sum_\mathbf{k} \abs{g_\mathbf{k}}^2\delta(\nu-\nu_\mathbf{k})$. 
In this paper, we consider spectral densities of the form:

\begin{equation}
    J(\nu) = \alpha \frac{\nu^s}{\nu_c^{s-1}}\mathrm{e}^{-\frac{\nu}{\nu_c}} \label{eq:spectral_density}
\end{equation}
with $\alpha$ denoting the overall coupling strength, $\nu_c$ the cutoff frequency of the environment, and $s$ the ohmicity or dimensionality of the environment. 

The resulting dynamics of the TLS are highly sensitive to the spectral character of the phonon environment. For instance, Ohmic environments are well known to exhibit a localized-to-delocalized quantum phase transition \cite{Silbey1983VariationalBath, Chin2006CoherentModel, Chin2011GeneralizedTransition, Tong2011QuantumModel}. Furthermore, for spectral exponents $s \leq 2$, the standard polaron transformation \cite{McCutcheon2011ADots}—typically employed in the spin-boson model to treat the tunneling rate $\Delta$ as the perturbation in a master equation formalism—becomes inapplicable. This is due to an infrared divergence in the associated Franck-Condon factor \cite{Silbey1983VariationalBath,Chassagneux2018EffectRegime}. To circumvent this limitation, Silbey and Harris \cite{Silbey1983VariationalBath} developed a variational transformation, which we introduce in the following section. Throughout this paper, we adopt the parameters $\alpha = 1 \Delta / (\nu_c \Gamma(s) )$ and $\nu_c = 10 \Delta$, with the bath temperature fixed at $k_B T = \hbar \Delta$. For these parameters, the reorganization energy $\int d\nu J(\nu) /\nu = \alpha \nu_c \Gamma(s)$ is fixed independent of the dimensionality $s$, with $\Gamma(s) = \int_0^\infty dx x^{s-1} \exp(-x)$

\subsection{The variational polaron transformation}
To go beyond the weak-coupling regime,  we utilize the variational polaron theory, in which a unitary transformation of the form 
\begin{equation}
    U_V = \dyad{1} \tilde{B}_- + \dyad{0} \tilde{B}_+
\end{equation}
where we defined the displacement bath operator
\begin{equation}
    \tilde{B}_\pm = \exp\left( \pm\sum_\mathbf{k} \frac{f_\mathbf{k}}{\nu_\mathbf{k}}(b_\mathbf{k}^\dagger - b_\mathbf{k})  \right),
\end{equation}
and $\{f_\mathbf{k}\}$ are a set of variational parameters, chosen to minimize the free energy of the open quantum system\cite{Silbey1983VariationalBath}. 
%
%
The effect of the transformation is thus to displace the bosonic bath either positively or negatively with $\tilde{B}_\pm$ depending on the state of the TLS. 

Performing the unitary transformation, the resulting Hamiltonian $H_V = U_V H U_V^\dagger$ reads:

\begin{align}
\begin{split}
{H}_{\rm V} &= \frac{\Delta_R}{2} \sigma_x + \sum_\mathbf{k} \nu_\mathbf{k} b_\mathbf{k}^\dagger b_\mathbf{k} \\
&+ \frac{\Delta}{2}(B_X \sigma_x + B_Y \sigma_y) + B_Z \sigma_z \label{eq:HV} 
\end{split}
\end{align}
where we defined the renormalized coupling rate $\Delta_R = \expval{B} \Delta $, with $\expval{B} = \expval{B_\pm}$ being the well-known Franck-Condon factor, and $B_\pm \equiv \tilde B_\pm^2$. 
The interaction between the bath and the TLS is defined in terms of the unitary operators
$B_X=(B_++B_--2\ev{B})/2$,  $B_Y=i(B_+-B_-)/2$, and  $B_Z = \sum_\mathbf{k} (g_\mathbf{k}-f_\mathbf{k})(b_\mathbf{k}^\dagger + b_\mathbf{k})$.

To determine the variational parameter $f_\mathbf{k}$, we follow Refs. \cite{McCutcheon2011ADots,Nazir2016ModellingDots,Denning2020OpticalInteractions} and minimize the Feynman-Bogoliubov bound on the free energy, 
yielding $f_\mathbf{k} = F(\nu_\mathbf{k})g_\mathbf{k}$, where 
%
\begin{align}
F(\nu_\mathbf{k}) &= \frac{1}{1 + \frac{2\Delta_\mathrm{R}}{\nu_\mathbf{k}} \tanh(\beta\hbar \Delta_\mathrm{R} /2) \coth(\beta\hbar\nu_\mathbf{k}/2)]}\label{eq:var_param}.
\end{align}
In the continuum limit, this yields the Franck-Condon factor,
\begin{equation}
    \expval{B} =\exp\left[-2 \int\limits_0^\infty \dd{\nu}\frac{J(\nu)F^2(\nu)}{\nu^2}\coth(\frac{\beta\hbar\nu}{2})\right].
\label{eq:R_B}
\end{equation}
While there are some circumstances in which analytic solutions exist \cite{Silbey1983VariationalBath}, in general, in order to specify the variational parameters, we must solve Eqs.~\ref{eq:var_param} and ~\ref{eq:R_B} self-consistently in a numerical fashion.

In the limit $F(\nu_\textbf{k}) \rightarrow 0$, we restore the original Hamiltonian, which leads to the weak master equation, because the transformation $U_{\rm V}$ reduces to the identity. In contrast, the limit $F(\nu_\textbf{k}) \rightarrow 1$ leads to the standard polaron transformation and its corresponding master equation. From the expression in Eq.~\eqref{eq:R_B}, it is evident that for $s\leq2$ and $F(\nu_\mathbf{k})=1$, the integral diverges leading to $\expval{B}=0$ for all bath parameters, which is a well-known shortcoming of the polaron transformation predicting incoherent dynamics regardless of the parameter regime \cite{Silbey1983VariationalBath,Harris1985VariationalSystem,Leggett1987Dynamics, Silbey1989TunnelingDescription,Chassagneux2018EffectRegime}. This was the original motivation for introducing the variational transformation~\cite{Silbey1983VariationalBath}, which overcomes this divergence by decoupling the lowest frequency modes, i.e., setting $F(\nu_\mathbf{k})\rightarrow 0 $ for $\nu \rightarrow 0$. 

\begin{figure}[!ht]
    \centering
    \includegraphics[width=\linewidth]{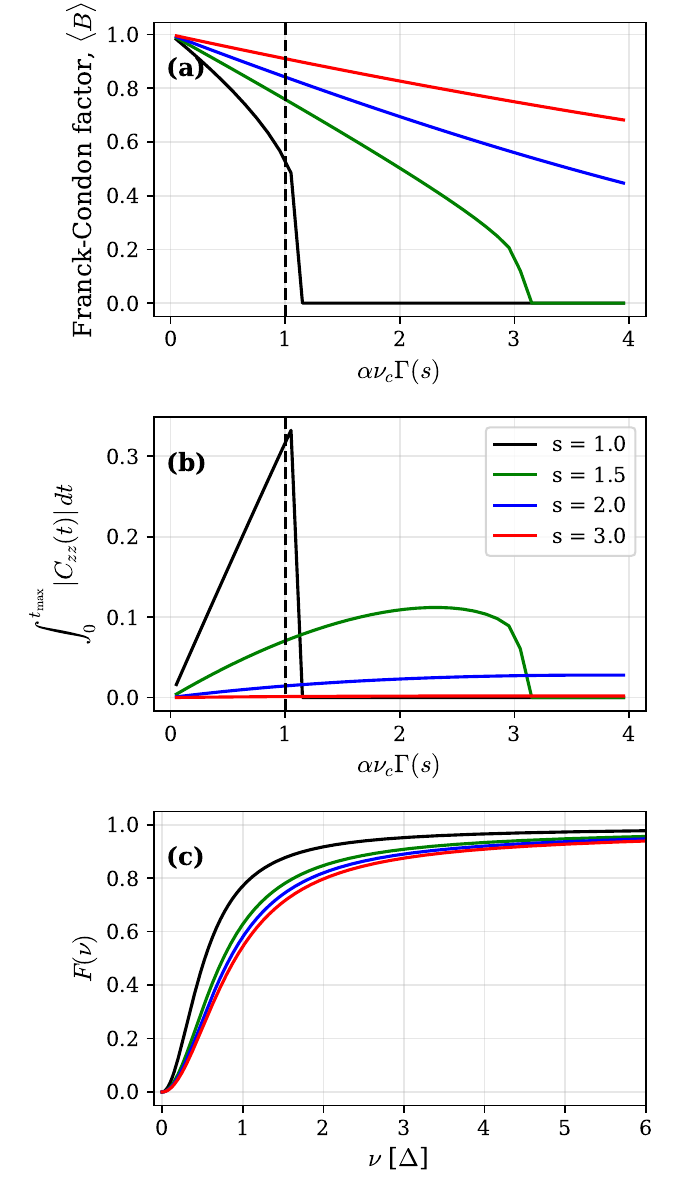}
    \caption{(a) The Franck-Condon factor $\expval{B}$, see Eq.~\eqref{eq:R_B}, shown as a function of the renormalization energy $\alpha \nu_c \Gamma(s)$, obtained by varying the dimensionless coupling $\alpha$, see Eq.~\eqref{eq:spectral_density} at fixed cutoff frequency $\nu_c$, for different dimensionalities $s$. For $s=1$ and $s=1.5$, we see a sudden transition where $\expval{B}  =0$ and the variational transformation no longer yields an appropriate basis for a master equation. (b): Same as (a), but the absolute integral of the correlation function $C_{zz}(t)$ as a function of the renormalization energy $\alpha \nu_c \Gamma(s)$. This shows that for lower dimensions, the slower modes, which are not transformed in the variational frame, constitute a larger contribution to the interaction. (c): The variational function $F(\nu)$ as a function of the frequency $\nu$ for $\alpha=0.1$ (indicated by the dashed line in (a)). We see that the low-frequency phonon modes are not included in the transformation $F(\nu)<1$.  }
    \label{fig:fracnk}
\end{figure}

\subsection{Restoring the Franck-Condon Factor}
The Franck-Condon factor is a crucial parameter for characterizing both optically active molecules and solid-state emitters. It naturally emerges from polaron theory\cite{Iles-Smith2017LimitsEmitters,Iles-Smith2017PhononSources} in the presence of super-Ohmic environments. 
For lower Ohmicities, the infrared divergence in Eq.~\ref{eq:R_B} means that $\langle B\rangle\rightarrow0$, and therefore polaron theory cannot capture, for example, non-Markovian sideband effects in emission spectra \cite{Iles-Smith2017LimitsEmitters,Denning2020PhononSources,Denning2020OpticalInteractions,Bundgaard-Nielsen2021Non-MarkovianElectrodynamics}, and the associated Franck-Condon factor, in these regimes.
The variational theory, however, enables us to extend the definition of the Franck-Condon factor to sub-Ohmic environments.

Figure \ref{fig:fracnk}(a) illustrates the Franck-Condon factor $B$ from Eq. \eqref{eq:R_B} across four dimensionality parameters $s \in \{1, 1.5, 2, 3\}$, with $\nu_c = 10 \Delta$ and $k_B T = \hbar \Delta$. 
For the Ohmic case ($s=1$), the factor drops discontinuously to zero at $\alpha \approx 0.1 \Delta$. This sudden transition signals a breakdown in the variational description, where the transformation fails to provide a physically meaningful ground-state ansatz. This behavior is a known signature of a phase transition toward effective localization \cite{Silbey1983VariationalBath, Chin2006CoherentModel, Chin2011GeneralizedTransition, Tong2011QuantumModel}, and is a consequence of the low-frequency modes becoming dominant in the behavior of the system. We note that using an asymmetric extension of the variational ansatz \cite{Chin2011GeneralizedTransition}, one can achieve an accurate description beyond this phase transition for $T=0$. However, for the unbiased spin-boson model at finite temperatures, the asymmetric extension of the variational transformation is equivalent to the variational transformation considered here.

Notably, the bath's influence is significant even prior to this jump, as evidenced by the diminished Franck-Condon factor ($\approx 0.5$).
While $s=1.5$ exhibits a similar discontinuity at higher coupling strengths, the super-Ohmic cases ($s=2, 3$) show no such transition and maintain generally higher Franck-Condon factors, indicating increased stability against environmental localization.

In Fig.~\ref{fig:fracnk}(b), we show the absolute integral over the correlation function $C_{ZZ}(t) = \expval{B_Z(t)B_z(0)}$, which provides a measure of the residual system–bath interaction that remains untransformed after applying the variational polaron transformation. As we show in Fig.~\ref{fig:fracnk}(c) and discuss shortly, the untransformed part of the Hamiltonian is mainly the slow, low-frequency modes of the bath. Thus, we also see that the magnitude of the correlation function $C_{ZZ}(t)$ is much larger for lower ohmicities, since they have a stronger influence of low-frequency modes. Again, the sudden kinks in the integral for $s=1$ and $s=1.5$ occur because the variational transformation no longer leads to a valid transformation.

To provide intuition for the variational approach, Fig.~\ref{fig:fracnk}(c) displays the variational function $F(\nu)$ from Eq.~\eqref{eq:var_param} at $\alpha = 0.1 \Delta$ [dashed line in Fig. \ref{fig:fracnk}(a)]. The function increases monotonically from zero at low frequencies to unity at high frequencies, marking the transition from undisplaced phonon modes to full polaron displacement. This behavior illustrates that low-frequency modes are "too slow" to respond to the system's characteristic transition rate ($\approx \Delta$); they remain untransformed and are excluded from the variational displacement. Because these low-frequency modes are subsequently treated perturbatively, strong coupling to these degrees of freedom can cause the perturbative expansion to fail, resulting in inaccurate dynamics. We discuss these implications further in Sec. \ref{sec:populations}.

\subsection{The variational master equation and initial state correlations\label{sec:var_me}}

While the projection operator formalism provides a systematic route to deriving master equations, it introduces a critical ambiguity regarding the precise choice of the reference environmental state, $\tau_R$. In standard derivations of the variational master equation, $\tau_R$ is typically chosen to be the thermal equilibrium state of the displaced bath \cite{McCutcheon2011ADots,Hsieh2019ATransport,Wang2020VariationalCrystals,Denning2020OpticalInteractions,Bundgaard-Nielsen2021Non-MarkovianElectrodynamics}. However, because this state is defined in the variational polaron frame, it does not correspond to a simple Gibbs state in the laboratory frame. 

Consequently, this choice imposes a non-trivial, implicit assumption about the initial system–bath correlations. To illustrate this, consider a standard uncorrelated initial state in the laboratory frame, $\chi(0) = \ket{1}\bra{1} \otimes \tau_B$. Applying the orthogonal projection operator $\mathcal{Q} = 1 - \mathcal{P}$ to this state in the variational frame yields:
\begin{equation}
\begin{aligned}
    \mathcal{Q} \chi(0) &= (1  - \mathcal{P}) (U_{\rm V} \ket{1}\bra{1} \otimes \tau_B U_{\rm V}^\dagger) \\
    &= \ket{1}\bra{1} \otimes (\tilde B_- \tau_B \tilde B_+ - \tau_R) \label{eq:Qchi0}
\end{aligned}    
\end{equation}
Consequently, neglecting the inhomogeneous terms in Eq. \eqref{eq:me_inhomo} (i.e., setting $\mathcal{Q} \chi(t_0)=0$) requires that $\tau_B = B_+ \tau_R B_-$. Physically, this implies that the initial bath state in the laboratory frame is not in simple thermal equilibrium, but is instead a displaced thermal state already correlated with the system. Conversely, to correctly model a completely uncorrelated, thermalized initial bath in the laboratory frame ($\tau_B = \tau_R$), the projection yields a non-vanishing contribution: $\mathcal{Q} \chi(t_0) = \ket{1}\bra{1} \otimes (B_- \tau_R B_+ - \tau_R)$. In this standard scenario, explicitly retaining the inhomogeneous terms is strictly necessary to accurately capture the ensuing dynamics as system–environment correlations build up.
In the following sections, we systematically investigate how these contrasting initial-state assumptions impact the dynamics and absorption spectrum of the TLS.

To derive the variational master equation, we partition the Hamiltonian in Eq.~\eqref{eq:Hamiltonian_split} into system operators, $A_i \in \{\frac{\Delta}{2}\sigma_x, \frac{\Delta}{2}\sigma_y, \sigma_z\}$, and corresponding environmental operators, $B_i \in \{B_X, B_Y, B_Z\}$. This specific decomposition yields five non-zero homogeneous bath correlation functions: $C_{XX}(\tau)$, $C_{YY}(\tau)$, $C_{ZZ}(\tau)$, $C_{YZ}(\tau)$, and $C_{ZY}(\tau)$, with all remaining cross-correlations vanishing. Explicit expressions for these terms are provided in Appendix~\ref{app:env_corr}. Furthermore, incorporating the inhomogeneous terms requires evaluating expectation values of the form $C^\mathcal{I}_{ij}(t,s) = \tr_\mathrm{B} [ B_{i}(t) B_j(s)(B_- \tau_R B_+ - \tau_R) ]$. As demonstrated in Ref.~\citenum{McCutcheon2011ConsistentEquation}, this expression can be conveniently recast using displaced bath operators, $B^V_i(t) = B_- B_i(t) B_+$. This transformation yields the simplified identity $C^\mathcal{I}_{ij}(t,s) = \tr_\mathrm{B} [ B^V_{i}(t) B^V_j(s) \tau_R ] - C_{ij}(t-s)$. The comprehensive evaluation of all such inhomogeneous correlation functions is detailed in Appendix~\ref{app:correlation_functions}.

\section{Expectation values in the polaron frame \label{sec:corrections}}

To evaluate laboratory-frame expectation values using the reduced density matrix obtained from the variational polaron master equation, the system observables must be mapped back to the original frame. As demonstrated in Ref.~\citenum{Iles-Smith2024CapturingFormalism}, system operators that do not commute with the variational polaron transformation become dressed by environmental degrees of freedom. Consequently, evaluating these observables requires explicit correction terms to account for this system–bath mixing. As we demonstrate below, critical corrections emerge naturally from the projection operator formalism.

The variational and lab frames are related as:
\begin{equation}
    \chi(t) = U_V \chi_L(t) U_V^\dagger
\end{equation}
where $\chi_L(t)$ is the density operator in the lab frame and $\chi(t)$ the density operator in the variational polaron frame, which is the frame in which we derive the master equation. Thus, the expectation value of some operator in the lab frame $A_L$ is given as:
\begin{equation}
    \expval{A_L(t)} =  \mathrm{tr}_{S+B}\left (A_L U_V^\dagger \chi(t) U_V \right) = \mathrm{tr}_{S+B}\left ( A_V  \chi(t) \right ) 
\end{equation}
where we defined the operator in the variational frame $A_V  = U_V A_L U_V^\dagger$. From the ME, we only have access to the system density matrix, and we thus use the projectors $I = \mathcal{P} + \mathcal{Q}$ to get:
\begin{equation}
     \expval{A_L(t)} = \expval{A_V(t)}_\mathrm{rel}  + \expval{A_V(t)}_\mathrm{irrel} \label{eq:rel_plus_irrel}
\end{equation}
where $\expval{A_V(t)}_\mathrm{rel} = \mathrm{tr}_{S+B}(A_V \mathcal{P} \chi(t))$ and  $\expval{A_V(t)}_\mathrm{irrel} = \mathrm{tr}_{S+B}(A_V \mathcal{Q} \chi(t))$. If $A_V$ acts exclusively on the system subspace, its expectation value over the irrelevant part of the density matrix vanishes by construction, yielding $\expval{A_V(t)}_\mathrm{irrel} = 0$. Conversely, if the original laboratory-frame operator $A_L$ does not commute with the variational polaron transformation, the transformed operator $A_V$ acquires a non-trivial bath dependence. This environmental dressing inevitably couples the observable to the irrelevant dynamics, resulting in a non-zero correction. 

Generally, any operator in the variational polaron frame can be expressed in the separable form:
\begin{equation}
A_V = \sum_\alpha s_\alpha \otimes B_\alpha,
\end{equation}
with $s_\alpha$ acting on the system degrees of freedom and $B_\alpha$ acting on the bath\footnote{
Notice that in all the following derivations, we reserve Greek letters in indices for operators associated with the expectation value operator $A_V$, whereas Latin indices are reserved for operators related to the propagators as in Sec.~\ref{sec:me}.}.
Evaluating the expectation value over the relevant part of the density matrix gives the familiar factorized form \cite{Nazir2016ModellingDots,Iles-Smith2024CapturingFormalism},
\begin{equation}
    \expval{A_V(t)}_\mathrm{rel} = \sum_\alpha \expval{s_\alpha(t)}\expval{B_\alpha(t)} \label{eq:rel_part}
\end{equation}
The corresponding irrelevant contribution was previously derived under the assumption of an initially uncorrelated variational state, $\mathcal{Q} \chi(t_0) = 0$ \cite{Iles-Smith2024CapturingFormalism}. In the present work, we lift this restriction to incorporate the inhomogeneous dynamics arising from $\mathcal{Q} \chi(t_0) \neq 0$. Inserting the equation of motion for the irrelevant projection [Eq.~\eqref{eq:q_eom}] leads directly to the complete expression (detailed in Appendix \ref{app:onetime}):
\begin{equation}
\begin{aligned}
    &\expval{A_V(t)}_\mathrm{irrel} = \sum _\alpha \Gamma_\alpha(t) \trS { s_\alpha(t) \rho_\mathrm{S}(t_0) } \\ 
    & -i \sum_\alpha \operatorname{tr}_\mathrm{S}\left[ s_\alpha(t) (
    \Psi_\alpha^\mathcal{I}(t)\rho_\mathrm{S} (t_0)+\Psi_\alpha(t) \rho_\mathrm{S} (t)\right.  \\
    &\left.- \rho_\mathrm{S}(t)\Theta_\alpha(t) + \rho_\mathrm{S}(t_0) \Theta_\alpha^\mathcal{I}(t)))\right],
    \label{eq:correction_terms}
\end{aligned}
\end{equation}
where we defined:
\begin{equation}
    \Psi_\alpha(t) = \sum_j \int_{0}^t \mathrm{~d} s C_{\alpha j}(t,s) A_j(s) , \label{eq:psi_alpha}
\end{equation}
\begin{equation}
    \Theta_\alpha(t) = \sum_j \int_{0}^t \mathrm{~d} s   C_{j \alpha }(s,t) A_j(s), \label{eq:theta_alpha}
\end{equation}
\begin{equation}
    \Psi_\alpha^\mathcal{I}(t) =  - \Phi_\alpha(t) + \sum_j \int_{0}^t \mathrm{~d} s C^\mathcal{I}_{\alpha j}(t,s) A_j(s) ,  \label{eq:psi_I_alpha}
\end{equation}
\begin{equation}
    \Theta_\alpha^\mathcal{I}(t) = - \Phi_\alpha(t) +\sum_j \int_{0}^t \mathrm{~d} s C^\mathcal{I}_{j\alpha }(s,t) A_j(s), \label{eq:theta_I_alpha}
\end{equation}
\begin{equation}
    \Phi_\alpha(t) = \expval{B_\alpha(t)} \sum_j \int_0^t ds \Gamma_j(t) A_j(t), \label{eq:phi_alpha}
\end{equation}
As emphasized previously, if the transformed operator $A_V(t)$ acts exclusively on the system subspace, the irrelevant correction terms strictly vanish. An environmental dependence is only acquired when the original laboratory-frame operator fails to commute with the variational polaron transformation. Consequently, population observables, such as $\sigma_z$, commute with the transformation and thus require no correction. In contrast, for off-diagonal observables like the coherence $\sigma_x$ and the corresponding absorption spectra, the transformation induces system--bath mixing, rendering these corrections essential. In the following, we derive the explicit forms of these non-vanishing contributions.

\subsection{Corrections to system coherences}
For a TLS, the transition operator $\sigma(t)$ transforms into the variational polaron frame as $\sigma_V(t) = \sigma(t) B_+(t)$. Consequently, physical observables that depend on these transitions, such as the dipole operator $\sigma_x(t) = \sigma(t) + \sigma^\dagger(t)$, are explicitly dressed by the bath and are therefore subject to the aforementioned corrections. Evaluating the relevant projection yields $\expval{\sigma_V(t)}_\mathrm{rel} = B \expval{\sigma(t)}_V$, where $\expval{\cdot}_V$ denotes the expectation value taken with respect to the reduced density matrix in the polaron frame, and $B$ represents the corresponding bath expectation value.

To evaluate the irrelevant contribution, $\expval{\sigma_V(t)}_\mathrm{irrel}$, we identify $B_\alpha = B_+$ in Eq.~\eqref{eq:correction_terms}. This requires computing both the homogeneous bath correlation functions, $C_{+i}(t,s) = \expval{B_+(t) B_i(s)} = C_{+i}(t-s)$, and their inhomogeneous counterparts, $C_{+i}^\mathcal{I}(t,s)$. The explicit derivation of these functions is provided in Appendix \ref{app:correlation_functions}. While the homogeneous term $C_{+i}(t-s)$ is time-translationally invariant and depends strictly on the time difference, the inhomogeneous term $C_{+i}^\mathcal{I}(t,s)$ retains a dependence on both absolute times $t$ and $s$ due to the initial system–bath correlations, and thus vanishes in the steady-state limit. Combining these relevant and irrelevant contributions, the full laboratory-frame expectation value of the coherence is given by:
\begin{equation}
\expval{\sigma_x(t)} = 2 \mathrm{Re}\left[ \expval{\sigma_V(t)}_\mathrm{rel} + \expval{\sigma_V(t)}_\mathrm{irrel} \right].
\end{equation}
  
\subsection{Corrections to the linear-response spectrum \label{sec:linear_correction}}
Beyond single-time expectation values, environmental dressing fundamentally alters two-time correlation functions. Consequently, these correction terms play a critical role in accurately calculating the linear-response absorption spectrum. To evaluate this spectrum, we adopt a standard perturbative approach wherein a weak external field $E(t)$ couples to the TLS \cite{Fetherolf2017LinearEquations,Svendsen2023SignaturesMaterials}. In the setting of the spin-boson model, this corresponds to a perturbation in the tunneling parameter $\Delta$. The total time-dependent Hamiltonian is given by$$H'(t) = H + E(t)\mu,$$where $\mu = \sigma_x = \sigma + \sigma^\dagger$ is the dipole transition operator of the system. In the case of the coherently driven quantum emitter, this can be experimentally achieved by tuning the power of the coherently driving laser or by probing with an additional laser pulse. In the weak-driving limit, one can apply time-dependent perturbation theory to the density matrix \cite{mukamel1995principles}. Assuming an impulsive excitation of the form $E(t) = \mathrm{e}^{i \omega t} \delta(t)$, this formalism leads directly to:
\begin{equation}
    A(\omega)= \lim\limits_{t \to \infty} 2 \operatorname{Re}\left[\int_0^{\infty} d \tau e^{i \omega \tau} S^{(1)}(t,\tau)\right], \label{eq:linear_response}
\end{equation}
where $S^{(1)}(t,\tau) = -i \mathrm{Tr}_\mathrm{SB} \{ \sigma_x^\dagger(t+\tau) [\sigma_x(t),\chi(0)] \} = \mathrm{Im} \left( \mathrm{Tr}_\mathrm{SB}\{ \sigma_x^\dagger(t+\tau) \sigma_x(t) \chi(0) \} \right)$ defines the linear response function. Here, $\chi(0) = \rho_\mathrm{S}(0) \otimes \tau_B$ represents the initial state of the composite system. In practice, the precise form of this initial preparation is inconsequential; we evaluate the response in the long-time limit ($t \to \infty$), ensuring the system has relaxed to its steady-state equilibrium prior to the impulsive excitation. Because $S^{(1)}(t,\tau)$ constitutes a two-time expectation value, its time evolution is governed by the modified quantum regression theorem derived in Sec.~\ref{sec:me_for_two_time}. 
Therefore, in the polaron frame:
\begin{equation}
    S^{(1)}(t,\tau) = \mathrm{Im}\!\left( \trSB{\sigma_{x,V}^\dagger(\tau)\tilde{\Lambda}(t,\tau)} \right),
\end{equation}
where $\tilde{\Lambda}(t,\tau)$ evolves according to Eq.~\eqref{eq:QRT} with initial condition
\begin{equation}
    \Lambda(t,0)=\sigma_{x,V}(\mathcal{P}+\mathcal{Q})\chi(t), \label{eq:lambda0}
\end{equation}
and
\begin{equation}
    \sigma_{x,V}=\sigma_{x,V}^\dagger=\sigma B_+ + \sigma^\dagger B_- = \sum_\alpha s_\alpha \otimes B_\alpha.
\end{equation}

The structure of $\mathcal{P}\Lambda(t,0)$ is then analogous to the decomposition used for the single-time expectation values in Eqs.~\eqref{eq:rel_part} and \eqref{eq:correction_terms}. Indeed, since
\begin{equation}
    \expval{\sigma_{x,V}} = \trSB{\sigma_{x,V}\mathcal{P}\chi(t)} + \trSB{\sigma_{x,V}\mathcal{Q}\chi(t)},
\end{equation}
we obtain
\begin{equation}
    \mathcal{P}\Lambda(t,0)
    =
    \trB{\sigma_{x,V}\mathcal{P}\chi(t)}\otimes\tau_R
    +
    \trB{\sigma_{x,V}\mathcal{Q}\chi(t)}\otimes\tau_R.
\end{equation}
The initial state for the QRT, therefore, already carries the influence of both the factorized part of the state and the system--environment correlations accumulated up to time $t$.
Inserting the explicit expressions for $\trB{\sigma_{x,V}\mathcal{P}\chi(t)}$ and
$\trB{\sigma_{x,V}\mathcal{Q}\chi(t)}$ yields
\begin{equation}
\begin{aligned}
\mathcal{P}\Lambda(t,0)
&=
\sum_\alpha \expval{B_\alpha}\,
s_\alpha(t)\tilde{\rho}_S(t)\otimes\tau_R \\
&\quad
-i \sum_\alpha s_\alpha(t)(\Psi_\alpha(t)\tilde{\rho}_S(t)
+
\tilde{\rho}_S(t)\Theta_\alpha(t)),
\end{aligned}
\label{eq:P_lambda_t0}
\end{equation}
where the first term corresponds to the factorized contribution, while the
remaining terms arise from the system--environment correlations contained in
$\mathcal{Q}\chi(t)$.

Here, we have kept only the contributions that survive in the limit
$t\rightarrow\infty$, since the linear-response spectrum is evaluated after the
system has relaxed to its steady state prior to the impulsive excitation.
Consequently, all terms depending explicitly on the initial state
$\rho(0)$ vanish. The full expression for an arbitrary $t$ is provided in
Appendix~\ref{app:QRT}.

As discussed in Sec.~\ref{sec:me_for_two_time}, the evolution of $\mathcal{P}\tilde{\Lambda}(t,\tau)$ contains an inhomogeneous contribution of the form $\mathcal{I}(\tau,0)[\mathcal{Q}\Lambda(t,0)]$ (Eq.~\eqref{eq:QRT}). Applying the projector $\mathcal{Q}$ to Eq.~\eqref{eq:lambda0} gives
\begin{equation}
\mathcal{Q}\Lambda(t,0)
=
\mathcal{Q}\sigma_{x,V}(t)\mathcal{P}\chi(t)
+
\mathcal{Q}\sigma_{x,V}(t)\mathcal{Q}\chi(t).
\end{equation}
The inhomogeneous contribution appearing in
Eq.~\eqref{eq:QRT} then becomes
\begin{equation}
\begin{aligned}
\mathcal{I}(\tau,0)[\mathcal{Q}\Lambda(t,0)]
&=
\mathcal{I}(\tau,0)[\mathcal{Q}\sigma_{x,V}(t)\mathcal{P}\chi(t)] \\
&\quad +
\mathcal{I}(\tau,0)[\mathcal{Q}\sigma_{x,V}(t)\mathcal{Q}\chi(t)] .
\end{aligned}
\label{eq:QRE_I}
\end{equation}
Here, $\tau$ and the integration variable $s\in[0,\tau]$ refer to the 
incremental propagation after the action of $\sigma_{x,V}(t)$, while the 
explicit dependence on $t$ keeps track of the state and correlations generated 
during the initial evolution from $0$ to $t$.

The first contribution arises from $\mathcal{Q}\sigma_{x,V}(t)\mathcal{P}\chi(t)$
and takes the form (see Appendix~\ref{app:QRT}).
\begin{equation}
\begin{aligned}
&I(\tau,0)[\mathcal{Q}\sigma_{x,V}(t)\mathcal{P}\chi(t)] = \\
& -i \sum_{\alpha i}\Gamma_{i\alpha}(\tau,t)
[A_i(\tau), s_\alpha(t)\rho_S(t)] \\
& - \sum_{ij\alpha}\int_{0}^{\tau} ds
\Big\{
C^{\mathcal I}_{ij\alpha}(\tau,s,t)
[A_i(\tau),A_j(s)s_\alpha(t)\rho_S(t)] \\
& \qquad +
C^{\mathcal I}_{ji\alpha}(s,\tau,t)
[s_\alpha(t)\rho_S(t)A_j(s),A_i(\tau)]
\Big\},
\end{aligned}
\label{eq:I_QP}
\end{equation}
where we defined $\Gamma_{j \alpha}(\tau,t) = \trB{B_j(\tau) (B_\alpha(t)\tau_R -\expval{B_\alpha(t)} \tau_R)}$ and $C^\mathcal{I}_{ij\alpha}(\tau,s,t) = C_{ij\alpha}(\tau,s,t)- \expval{B_\alpha(t)} C_{ij}(\tau,s)$ with $C_{ij\alpha}(\tau,s,t) = \trB{B_i(\tau)B_j(s)B_\alpha(t) \tau_R}$.

The second contribution originates from the correlated part
$\mathcal{Q}\sigma_{x,V}(t)\mathcal{Q}\chi(t)$ and, keeping terms up to
$\lambda^2$, yields (see Appendix~\ref{app:QRT})
\begin{equation}
\begin{aligned}
&I(\tau,0)[\mathcal{Q}\sigma_{x,V}(t)\mathcal{Q}\chi(t)] = \\
& - \sum_{ij\alpha}\int_{0}^{t} ds
\Big\{
C_{i\alpha j}(\tau,t,s)
[A_i(\tau),s_\alpha(t)A_j(s)\rho_S(t)] \\
& \qquad +
C_{ji\alpha}(s,\tau,t)
[s_\alpha(t)\rho_S(t)A_j(s),A_i(\tau)]
\Big\}.
\end{aligned}
\label{eq:I_QQ}
\end{equation}

We therefore see that the evaluation of the inhomogeneous contribution now
requires three-time bath correlation functions
$C_{ij\alpha}(\tau,s,t)$ and $C^{\mathcal I}_{ij\alpha}(\tau,s,t)$,
which are calculated explicitly in
Appendix~\ref{app:correlation_functions}.
With Eqs.~\eqref{eq:P_lambda_t0} and \eqref{eq:QRE_I} we can determine
$\Lambda(t,\tau)$ by evolving $\mathcal{P}\Lambda(t,0)$ according to
Eq.~\eqref{eq:QRT}.

As in the case of single-time observables, the final two-time correlation
function separates naturally into relevant and irrelevant contributions:
\begin{equation}
S^{(1)}(t,\tau)=S^{(1)}_{\mathrm{rel}}(t,\tau)+
S^{(1)}_{\mathrm{irrel}}(t,\tau).
\label{eq:linear_response_correction}
\end{equation}

The relevant part takes the form
\begin{equation}
\begin{aligned}
S^{(1)}_{\mathrm{rel}}(t,\tau)
&=\mathrm{tr}_{SB}
\{\sigma_{x,V}^\dagger(\tau)\mathcal{P}\tilde{\Lambda}(t,\tau)\} \\
&=\sum_\alpha
\langle B_\alpha(\tau)\rangle
\mathrm{tr}_S\{s_\alpha(\tau)\tilde{\rho}_S(t,\tau)\},
\end{aligned}
\end{equation}
where
\[
\tilde{\rho}_S(t,\tau)=
e^{iH_S\tau}\,
\mathrm{tr}_B\{\Lambda(t,\tau)\}\,
e^{-iH_S\tau}.
\]

The remaining contribution
\begin{equation}
S^{(1)}_{\mathrm{irrel}}(t,\tau)
=\mathrm{tr}_{SB}
\{\sigma_{x,V}^\dagger(\tau)\mathcal{Q}\tilde{\Lambda}(t,\tau)\} \label{eq:spec_irrel}
\end{equation}
originates from the irrelevant part of the system–environment correlations.
Its full expression is lengthy and is given in
Appendix~\ref{app:two_time}.

\subsection{Magnitude of corrections}

It is helpful to estimate the scale of the corrections introduced to better appreciate when they are necessary. In the following, we give the approximate magnitudes of these corrections. 

 For the corrections to the coherences $\expval{\sigma_x}$, we can estimate the magnitude of the correction as:
\begin{equation}
 \delta_x = \frac{\expval{\sigma_{x,V}(t)}_{\rm irrel}}{\expval{\sigma_x(t)}} \approx 2 \Delta B^2\int_0^\infty d\nu \frac{4J(\nu)F^2(\nu)}{\nu^3}.
\end{equation}
which follows from considering the steady state correction and thus the magnitude of the operators $\Psi_\alpha(t)$ and $\Theta_\alpha(t)$. Considering the $Y$ components, which has $C_{+Y} = -i\expval{B}^2 \sinh(\phi(\tau))$, the relevant magnitude is:
\begin{equation}
\begin{aligned}
    &2 \Delta \mathrm{Re} \left(\int_0^\infty \mathrm d\tau\,  C_{+Y}(\tau) \right) \\ 
    &\approx 2 \Delta B^2 \int \int d \tau d \nu \frac{4 J(\nu)F^2(\nu)}{\nu^2} \sin(\nu\tau)  
    \\
    &= 2 \Delta B^2 \int d \nu \frac{4 J(\nu)F^2(\nu)}{\nu^3}     
\end{aligned}
\end{equation}
where we used that $\int d \tau \sin(\nu \tau) = 1/\nu$ and estimated $\sinh(\phi(\tau)) \approx \phi(\tau)$

For $s>2$, we can set $F(\nu)=1$ and estimate it as:
\begin{equation}
\delta_x \approx \frac{8 \Delta B^2\alpha}{\nu_c}\Gamma(s-2).
\end{equation}
For $s<2$, the integral is dominated by the low-frequency part, and we approximate $F^2(\nu)$ by a soft step function cutting off frequencies below $\nu_F$
For $\nu_F \ll\nu_c$ and $s<2$, this gives
\begin{equation}
\begin{aligned}
    & \delta_x  \approx 2 \Delta B^2\int_0^\infty d\nu \frac{4J(\nu)F^2(\nu)}{\nu^3}
\approx
\frac{8 \Delta B^2\alpha}{\nu_c}
\int_{\nu_F/\nu_c}^{1} dx x^{s-3}
\\ 
&\approx
\frac{8 \Delta B^2\alpha}{(2-s)\nu_c}
\left(\frac{\nu_c}{\nu_F}\right)^{2-s}.    
\end{aligned}
\end{equation}
For $s=2$ one arrives instead at:
\begin{equation}
\delta_x \approx\frac{8 \Delta B^2\alpha}{\nu_c}
\ln\left(\frac{\nu_c}{\nu_F}\right).
\end{equation}
In the above, we defined $\nu_F$ as the point at which $F(\nu_F)=1/2$, which in the low-temperature limit, $\beta\nu_F\gg1$, gives
\begin{equation}
\nu_F
\approx
2\Delta_{\rm R}
\tanh\left(\frac{\beta\Delta_{\rm R}}{2}\right),
\end{equation}
while in the high-temperature limit, $\beta\nu_F\ll1$,
\begin{equation}
\nu_F
\approx
\sqrt{
\frac{4\Delta_{\rm R}}{\beta}
\tanh\left(\frac{\beta\Delta_{\rm R}}{2}\right)
}.
\end{equation}

Evaluating these estimates for the parameters used in this paper (see Sec.~\ref{sec:model_and_var}) gives the values presented in Table~\ref{tab:corrections}:
\begin{table}[htbp]
\centering
\caption{Comparison between the heuristic estimate for the relative correction to the steady-state coherence, $\delta_x$, and its exact value obtained from the full variational master equation, for the four ohmicities considered in this work. The estimate agrees well with the exact value except for $s=1$, where strong coupling to slow bath modes drives the perturbative expansion beyond its regime of validity. See Sec.~\ref{sec:coherences}) for the results showing the corrections.}
\label{tab:corrections}
\begin{tabular}{|c|c|c|c|c|}
\hline
\textbf{$s$} & \textbf{1} & \textbf{1.5} & \textbf{2} & \textbf{3} \\\hline
$\delta_x$ (estimate) & 0.30 & 0.32 & 0.12 & 0.03 \\\hline
$\delta_x$ (exact) & 1.5 & 0.31 & 0.11 & 0.029 \\\hline
\end{tabular}
\end{table}

Thus, we expect the corrections $\delta_x$ to be between $30\%-3\%$ of the steady state value. The actual values (which can also be seen in Sec.~\ref{sec:coherences}) are very close to the predicted corrections. Thus, the estimate is a good indicator of the importance of the correction terms. Note that the deviation for $s=1$ is due to the coupling being too strong for a second-order expansion to be accurate. 

The corrections to expectation values which occur due to the inhomogeneous terms will only be relevant on timescales shorter than the bath correlation time $\approx 1 / \nu_c$, since they capture the relaxation of the bath from its non-equilibrium initial state, which we confirm below in Sec.~\ref{sec:coherences}.

As for the corrections to the quantum regression theorem, these can be estimated through the correction to the dynamics given by: $\Gamma_{i\alpha}(\tau,t)$, which for $\tau=t$, and the $Y$ contribution, which has the largest magnitude: $\Delta/2 \abs{\Gamma_{i\alpha}(t,t)} = \Delta/4 (1-B^4)$. In the homogeneous dynamics, the approximate magnitude is roughly \cite{McCutcheon2010QuantumRegime} $\Delta^2/(8\nu_c) (1-B^4)$. We have $\Delta/\nu_c=1/10$ and thus, the additions to the quantum regression theorem are generally significant. In Sec.~\ref{sec:linear_response}, we see this manifest as large corrections to the linear response spectrum. This is even further illustrated in the case of the weak master equation, where the corrections are necessary even to predict a non-zero linear response.

\section{Single-time dynamics \label{sec:populations}}

\begin{figure}[!ht]
    \centering
    \includegraphics[width=\linewidth]{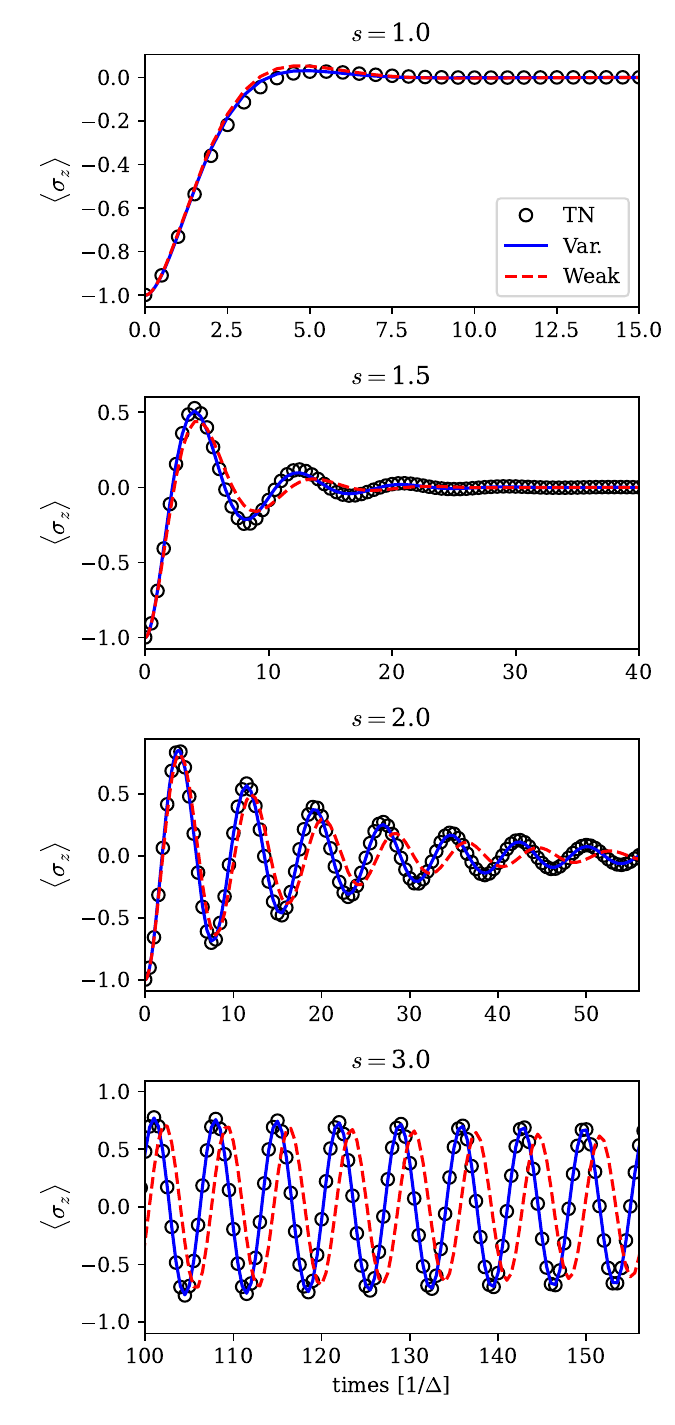}
    \caption{The population $\expval{\sigma_z}$ as a function of time for different ohmicities $s$ with the TLS initially in $\rho_S(0) = \dyad{1}$, as calculated using the variational ME (Var.), weak ME (Weak), and tensor network (TN). Parameters are $\alpha = \Delta / (\nu_c \Gamma(s))$, $\nu_c = 10 \Delta$, and a temperature of $k_B T = \hbar \Delta $. For the tensor network, we use $\epsilon = 5 \cdot 10^{-10}$, $\Delta t = 0.02/\Delta$, $T_{mem} = \{1000/\Delta,1000/\Delta,1000/\Delta,200/\Delta \}$ for the four ohmicities, respectively.}
    \label{fig:excited_false}
\end{figure}
In this section, we benchmark the accuracy of the variational transformation and our newly derived correction terms against numerically exact tensor-network simulations. For the tensor-network calculations, we employ the methodology and open-source implementation detailed in Ref.~\citenum{Link2024OpenContraction}. We provide additional numerical details in Appendix \ref{app:tensor_details}, as well as provide the numerical costs of the master equation simulations in Appendix \ref{app:numerical_costs}. See also Appendix \ref{app:error_curves}, where we plot the deviation between the master equation approaches and the tensor network calculations. As an additional baseline, we also evaluate the dynamics using the standard weak-coupling master equation, recovered within our formalism by setting $f_\mathbf{k} = 0$. In this untransformed limit, all system observables trivially commute with the identity transformation; consequently, neither environmental dressing corrections nor inhomogeneous terms arise. Crucially, this benchmarking framework allows us to explicitly test the impact of the inhomogeneous driving terms. By systematically comparing simulations with and without these terms for both population and coherence dynamics, we can precisely distinguish between an environment initially prepared in a standard thermal state versus one in a displaced thermal state.

\subsection{Populations}
We start by analyzing the evolution of the population difference, $\expval{\sigma_z(t)}$. The TLS is initialized in the state, $\rho_\mathrm{S}(0) = \ket{1}\bra{1}$. We find that explicitly including the inhomogeneous terms from Eq.~\eqref{eq:me_inhomo} produces no discernible change in the population dynamics, indicating that initial system--bath correlations do not significantly perturb the diagonal elements of the reduced density matrix. Consequently, we omit these terms for the population calculations. As we show in the following section, system coherences are far more sensitive to this initial environmental state.

In Fig.~\ref{fig:excited_false}, we compare the population dynamics obtained from the exact tensor-network method, the variational master equation, and the standard weak-coupling master equation across four Ohmicities: $s \in \{1, 1.5, 2, 3\}$. For the strictly Ohmic case ($s = 1$), both master equations agree remarkably well with the tensor-network benchmark. The success of the weak-coupling approach, which is perturbative in the system-bath coupling, is somewhat surprising given the strong system--bath interaction, which drives the rapid damping of the population oscillations.

However, as the spectral density becomes super-Ohmic, the limitations of the weak-coupling approximation emerge. For $s = 1.5$ and $s = 2$, its predictions begin to deviate noticeably from the exact dynamics, whereas the variational master equation maintains excellent accuracy. This performance gap is most pronounced in the super-Ohmic regime ($s = 3$), where the weak-coupling approach completely fails to capture the transient dynamics, while the variational method continues to yield robust, accurate agreement. It appears that for $s=1$, the dominance of slow, low-frequency modes, which do not rapidly dress the system, can be well captured by the weak-coupling master equation. For $s=3$, in contrast, the importance of fast phonon modes is accurately captured by the variational transformation, while remaining beyond the reach of the weak-coupling master equation.

\subsection{Coherences \label{sec:coherences}}

\begin{figure}[!ht]
    \centering
    \includegraphics[width=\linewidth]{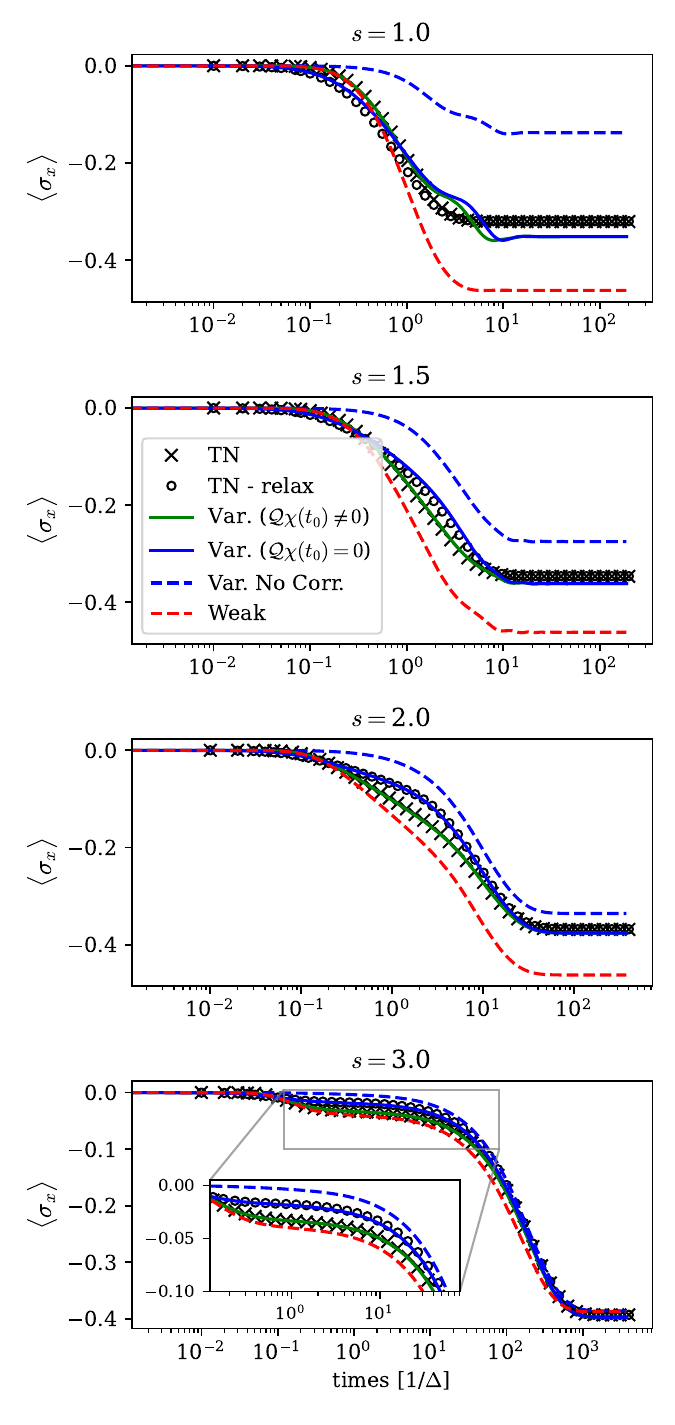}
    \caption{The coherences $\expval{\sigma}_x$ as a function of time for different ohmicities $s$ with the two-level system initially in $\rho_S(0) = \dyad{1}$ as calculated using the variational ME with corrections and inhomogeneous terms, with corrections but no inhomogeneous terms, and completely without corrections. We also show the weak ME and the tensor network with and without an initial relaxation of the bath. Parameters are the same as in Fig.~\ref{fig:excited_false}.}
    \label{fig:ground}
\end{figure}
In contrast to the populations, the dynamics of the coherence $\expval{\sigma_x}$ are highly sensitive to the initial state of the environment \cite{Trushechkin2019CalculationTransfer}. While the TLS is still initialized in the state $\rho_S(0) = \dyad{1}$, omitting the inhomogeneous terms ($\mathcal{Q} \chi(t_0)=0$) mathematically assumes the bath begins in a displaced thermal state, $\tau_B = B_- \tau_R B_+$ (see Eq.~\eqref{eq:Qchi0}).
Physically, this corresponds to the equilibrium configuration the bath naturally adopts when coupled to the TLS via the interaction Hamiltonian $H_I$. 
To instead model an environment that is initially uncoupled from the TLS (i.e., a laboratory-frame product state), we must incorporate the inhomogeneous contributions into the variational master equation ($\mathcal{Q} \chi(t_0)\neq0$).

The numerically exact tensor-network method, however, implicitly assumes that the bath is initially uncoupled from the system and prepared in a laboratory-frame Gibbs state. Therefore, a direct comparison requires incorporating the inhomogeneous terms into the variational master equation. An alternative approach is to artificially "relax" the environment within the tensor-network simulation. Specifically, we set the bare system Hamiltonian to zero ($H_\mathrm{S}=0$) and evolve the TLS initially in $\rho_S(0) = \dyad{1}$ for a duration $t_\mathrm{relax}$ strictly under the influence of the environment. This procedure leaves the reduced state of the system $\rho_S$ invariant while allowing the bath to naturally equilibrate into the exact displaced thermal state, $\tau_B = B_- \tau_R B_+$. Subsequent evolution under the full Hamiltonian ($H_\mathrm{S}\neq 0$) then captures the coherence dynamics corresponding to this displaced initial bath configuration. We can explicitly confirm this equivalence by comparing the dynamics of the "relaxed" tensor network against the variational master equation evaluated without inhomogeneous terms. For all numerical simulations, we utilize a relaxation time of $t_\mathrm{relax} = T_\mathrm{mem}$, which safely exceeds the characteristic memory time of the environment.

In Fig.~\ref{fig:ground}, we plot the time evolution of the coherences, $\expval{\sigma_x(t)}$, for the four Ohmic parameters $s \in \{1, 1.5, 2, 3\}$. We evaluate the variational master equation under two initial conditions: an initially displaced thermal bath ($\mathcal{Q}\chi(t_0)=0$) and an initially uncoupled, standard thermal bath ($\mathcal{Q}\chi(t_0) \neq 0$). Crucially, both of these calculations explicitly incorporate the environmental dressing corrections derived in Eq.~\eqref{eq:correction_terms}. These results are directly benchmarked against the exact tensor-network simulations, shown both with and without the artificial pre-relaxation step. For completeness, we also plot two baseline models: the "standard" variational master equation, which omits all inhomogeneous and correction terms entirely, and the untransformed weak-coupling master equation.

Beginning with the super-Ohmic regime ($s=3$), the impact of the initial bath state is immediately evident in the distinct curves predicted by the two tensor-network simulations, though both eventually converge to identical steady-state values. 
Furthermore, the variational master equation incorporating the inhomogeneous driving terms ($\mathcal{Q}\chi(t_0)\neq0$) accurately reproduces the dynamics of the unrelaxed tensor network, confirming that both methods correctly represent an initially uncoupled thermal bath. Conversely, omitting the inhomogeneous terms yields excellent agreement with the pre-relaxed tensor-network data. This confirms our theoretical prediction: without explicit inhomogeneous driving, the variational formalism inherently assumes the environment begins in the displaced thermal state.

Beyond the effects of the initial preparation, the standard weak-coupling master equation fails to accurately predict the coherence dynamics across all Ohmic regimes, underscoring the critical necessity of the variational polaron transformation. Equally importantly, we observe that the baseline variational theory—when evaluated without our newly derived correction terms—completely fails to capture the initial, short-time loss of coherence. This stark deviation highlights the essential role these dressing corrections play in properly capturing the exact early-time system–environment entanglement.

Moving to the intermediate regimes ($s=2$ and $s=1.5$), the variational master equation maintains excellent agreement with the exact tensor-network calculations. However, for the strictly Ohmic case ($s=1$), the variational approach struggles to reproduce the exact transient dynamics and incorrectly predicts the steady-state coherence. We attribute this breakdown to the dominant influence of slow, low-frequency phonon modes, which are not adequately captured by the variational transformation. As previously illustrated in Fig.~\ref{fig:fracnk}(b), the correlation function $C_{ZZ}(\tau)$—which represents the residual, untransformed system–environment coupling—exhibits a significantly larger magnitude in this Ohmic regime compared to the super-Ohmic cases. Consequently, the underlying second-order Born approximation with respect to this residual interaction Hamiltonian becomes insufficient. This perturbative breakdown is further evidenced by the exceptionally large magnitude of the dressing corrections themselves, visible as the stark difference between the corrected and uncorrected variational curves. To confirm this physical limitation, we demonstrate in Appendix~\ref{app:s1_weak} that reducing the system–bath coupling strength to $\alpha = 0.05 \Delta$ restores excellent agreement between the tensor-network and variational predictions for $s=1$.

\section{Linear response \label{sec:linear_response}}

\begin{figure}[!ht]
    \centering
    \includegraphics[width=\linewidth]{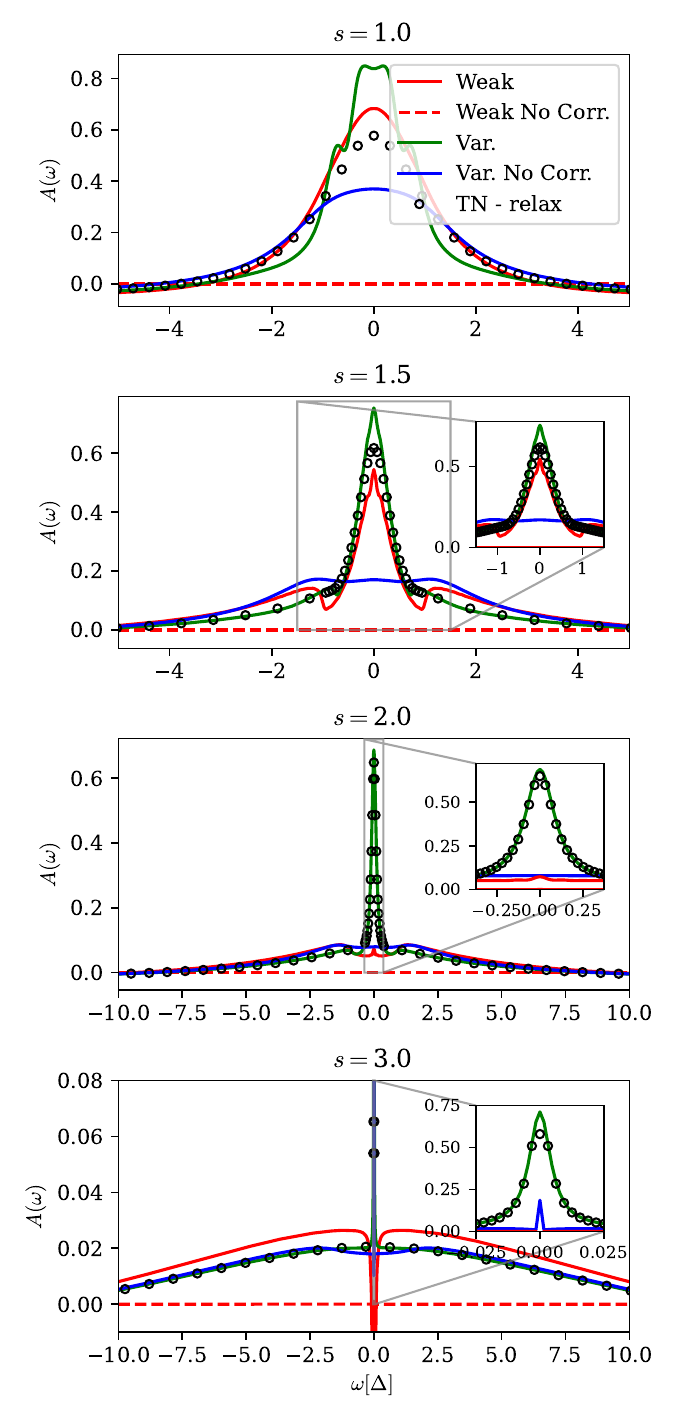}
    \caption{The linear response spectrum $A(\omega)$, see Eq.~\eqref{eq:linear_response} for different ohmicities $s$, as calculated using the Variational ME with the corrections in Eq.~\eqref{eq:linear_response_correction}, and using the typical formula without corrections in Eq.~\eqref{eq:seperating}. We also show the results using the Weak ME with the extended QRT (the additional driving term of Eq.~\eqref{eq:QRE_I}) and without the extension. Without the non-Markovian extension $A(\omega)=0$ for the Weak ME. For reference, we also show the linear response spectrum as calculated using the tensor network, where it has been relaxed to steady state and then the operator $\sigma_x$ has been applied, and then the state subsequently evolved.}
    \label{fig:linear_response}
\end{figure}

To evaluate the linear response via the quantum regression theorem, we first propagate the composite system to its steady state before applying the excitation operator, $\sigma_x$. As shown previously, incorporating the correlated system--environment component, $\mathcal{Q} \chi(t)$, is critical for accurately capturing single-time expectation values and steady-state dressing corrections. Extending this framework to two-time correlation functions requires explicitly tracking the action of the excitation operator on this correlated state, $\sigma_x(t) \mathcal{Q}\chi(t)$, throughout the subsequent time evolution. This comprehensive treatment contrasts sharply with standard literature approaches. Conventional calculations of spectra within the variational polaron frame typically neglect these higher-order inhomogeneous terms, assuming instead thermal equilibrium and system--environment separability \cite{Iles-Smith2017LimitsEmitters, Denning2020PhononSources,Roy2012PolaronSystem,Bundgaard-Nielsen2021Non-MarkovianElectrodynamics}:
\begin{equation}
\begin{aligned}
\expval{\sigma ^\dagger(t+\tau) \sigma(t)}_L &= \expval{\sigma ^\dagger(t+\tau) B_-(t+\tau) \sigma(t) B_+(t)}_V \\
&\approx \expval{B_-(t+\tau)B_+(t)}_V \expval{\sigma ^\dagger(t+\tau) \sigma(t)}_V.  \label{eq:seperating}
\end{aligned}
\end{equation}

This factorization inherently omits both the dynamical dressing corrections of Eq.~\eqref{eq:QRE_I} and the correlated inhomogeneous contributions of Eq.~\eqref{eq:spec_irrel}—an approximation that, as demonstrated below, severely compromises the accuracy of the predicted linear response.

Furthermore, we note that the inhomogeneous contribution to the QRT from $\sigma_x(t) \mathcal{Q} \chi(t)$ in Eq.~\eqref{eq:QRE_I} is non-vanishing even in the weak ME, see also Ref.~\cite{mccutcheon_optical16}. Below, we also show the effect of including this correction to the QRT.

For an exact benchmark, we compare these predictions against numerically exact tensor-network simulations. To prepare the requisite steady state prior to applying the excitation operator $\sigma_x$, we repeat the protocol from the preceding section: we evolve the TLS initially in $\rho_S(0)=\dyad{1}$ under the full Hamiltonian until equilibration is achieved, apply $\sigma_x$, propagate the composite state for a delay time $\tau$, and finally evaluate the expectation value of $\sigma_x$ to yield the two-time correlation function.

In Fig.~\ref{fig:linear_response}, we present the linear response spectra across the four Ohmic regimes (see also Appendix \ref{app:error_curves}, where we plot the deviation between the master equation approaches and the tensor network calculations). We compare the tensor-network benchmark against the variational master equation, including the correction terms derived in Sec.~\ref{sec:linear_correction}. For reference, we also plot the standard factorized approximation from Eq.~\eqref{eq:seperating}, the weak ME without the correction to the QRT, and with the correction to the QRT.

For a coherently driven TLS, the linear response in Eq.~\eqref{eq:linear_response} corresponds to an instantaneous perturbation to the driving strength $\Delta$ - either by a second laser field or by adjusting the power of the driving laser. If phonons were not present, we would have $\comm{H}{\sigma_x} = 0$ and this perturbation would therefore not couple the eigenstates of the system with each other, and no response is expected. In appendix \ref{app:linear_response}, we further show that with the introduction of a simple Markovian Lindblad term with the jump operator $\sigma_z$, the linear response is still vanishing.

Importantly, we see in Fig.~\ref{fig:linear_response}, that without the inclusion of the extended QRT, the weak ME predicts a linear response spectrum of $A(\omega)=0$. This is a clear indication that without the inclusion of non-Markovian corrections, the standard QRT fails to correctly predict the linear response spectrum. Furthermore, this also indicates that the linear response spectrum is an experimentally accessible way to probe the properties of the system's phononic environment. Importantly, the presence of a linear response indicates non-Markovianity in the coupling with the environment. 

Considering the linear response as predicted by the exact tensor network calculations, we generally see a central narrow peak for $s=2$ and $s=3$, which is consistent with the slower damping rate of these two environments, which we also saw in Fig.~\ref{fig:excited_false}, where for $s=2$ and $s=3$, the population undergoes multiple Rabi oscillations before dying out. In contrast, for $s=1.5$ and $s=1$, the damping is much stronger (as also seen in Fig.~\ref{fig:excited_false}), and correspondingly, the central peak in Fig.~\ref{fig:linear_response} broadens. For $s=2$ and $s=3$, we observe broad spectral features associated with the quick response of the phonon bath, while for $s=1$ and $s=1.5$, the dominating slow phonon modes make the broad spectral features less pronounced. 

Comparing now the tensor network calculations with the perturbative master equations, we see that for the weak ME with the extended QRT, we observe large deviations for $s=2$ and $s=3$, while for $s=1$ and $s=1.5$, the weak ME is close to the tensor network predictions. Especially for $s=1$, the weak ME seems to be more accurate than the variational ME, where we observe significant deviations between the tensor network and both variational approaches. As discussed in the previous section, this discrepancy stems from the strong residual coupling to slow phonon modes. However, as demonstrated in Appendix~\ref{app:s1_weak}, reducing the coupling strength to $\alpha=0.05 \Delta$ restores excellent agreement for the corrected variational method, whereas the uncorrected approach still exhibits substantial errors. Noticeably, the weak ME seems to be even more accurate than the variational approach, including correction terms. It thus seems that the variational transformation is not a helpful frame to derive the perturbative master equation from when $s=1$, where we generally see the weak ME perform well. Again, this is due to the strong effects of the slow phonon modes, which are not changed by the variational polaron transformation.  

For the super-Ohmic regimes, the corrected variational master equation performs exceptionally well, accurately reproducing both the central absorption peak and the phonon sideband predicted by the tensor network. Conversely, while the uncorrected variational approach captures the broad phonon sideband, it fails significantly to predict the correct amplitude and shape of the central peak. This illustrates the necessity of incorporating the non-equilibrium environmental state—and its corresponding dynamical corrections—when evaluating two-time correlation functions in the polaron frame.

\section{Outlook and Conclusion \label{sec:conclusion}}

In this work, we have derived and analyzed an extended quantum regression theorem for the non-Markovian perturbative variational polaron master equation that explicitly accounts for system–environment correlations and non-equilibrium bath states. Using the projection operator formalism, we included the inhomogeneous contributions that are typically neglected in standard applications of the quantum regression theorem and showed how these terms modify both single-time observables and two-time correlation functions. Within the variational polaron framework, these corrections arise naturally from the mixing of system and bath degrees of freedom induced by the transformation itself.

Focusing on the spin–boson model with spectral densities of varying ohmicity, corresponding to the dimension of the physical system for deformation potential interactions, including the ohmic case (1D), we demonstrated that these inhomogeneous contributions are essential for obtaining consistent results. In particular, parameter regimes where the standard polaron transformation exhibits infrared divergences can be treated using the variational approach. Benchmarking against numerically exact tensor-network simulations shows that the resulting theory accurately captures both transient dynamics and steady-state two-time correlations, including linear-response spectra, even at strong system–bath coupling. We do find, however, that the dominance of slow low-frequency modes in the ohmic case, which are unaffected by the variational transformation, necessitates lower system-bath couplings for accurate results.

Beyond quantitative agreement, the treatment provides a clear physical interpretation of the role of system--environment correlations in the dynamics, both those present in the initial state and those that build up during the evolution. The variational transformation provides an effective description of the bath through the renormalized coupling, yielding an intuitive picture of how environmental dressing modifies system dynamics. This perspective is especially valuable in regimes where the conventional polaron transformation breaks down, such as low-dimensional environments with ohmic or near-ohmic spectral densities. Examples of systems with such spectral densities include quantum emitters formed from defects in 2D materials \cite{Wang2016CoherentTemperature, Han2018RabiTemperature,Selig2016ExcitonicDichalcogenides,Fischer2023CombiningOrigin}, quantum dots embedded in 1D nanophotonic wires \cite{FerreiraNeto2024One-dimensionalDimensionality}, or trapped excitons in single carbon nanotubes (1D) \cite{Watahiki2012Enhancementnanocavities,Jeantet2016WidelyRegime,Jeantet2017ExploitingNanotube}.

Moreover, by explicitly identifying the role of the bath initial state, our framework enables a transparent comparison with numerically exact tensor-network methods, whose underlying assumptions about environmental preparation are often implicit.

While the present work has focused on the long-time limit of the quantum regression theorem, the formalism naturally extends to finite and arbitrary delay times. Exploring the full time-dependence of the inhomogeneous contributions is a natural direction for future work and would provide further insight into the buildup and decay of system–bath correlations, as well as into the temporal range of validity of the perturbative expansion.

Furthermore, all main results are derived using a general form of the Hamiltonian, and extensions to multi-site spin boson models are therefore straightforward. Such extensions already exist, although they do not include the corrections presented here \cite{Pollock2013}. For the extension with the corrections, new correlation functions will have to be computed.

\begin{acknowledgments}
We thank Ahsan Nazir and Jesper Mørk for insightful conversations. This work was supported by the Danish National Research Foundation through NanoPhoton - Center for Nanophotonics, Grant No. DNRF147.
\end{acknowledgments}

\clearpage
\appendix

\section{The projection operator formalism}

In this appendix, we provide additional derivations of some corrections and inhomogeneous evolutions that occur when the total state of the system and bath is included through the use of $\mathcal{P}$ and $\mathcal{Q}$ operators in the projection operator formalism.

\subsection{Corrections to one-time expectation values\label{app:onetime}}
Considering one-time expectation values $\expval{A_V(t)} = \trSB{A_V(t) \chi(t)}$ when the operator $A_V(t)$ is acting on both the environment and system parts leads to corrections from the often omitted $\mathcal{Q} \chi(t)$ part. We decompose the operator $A_V(t)$ as:
\begin{equation}
    A_V  = \sum_\alpha s_\alpha \otimes B_\alpha
\end{equation}
Now, we do not directly have access to $\chi(t)$, so we insert the identity $I =\mathcal{P} + \mathcal{Q}$, which leads to Eq.~\eqref{eq:rel_plus_irrel}. Taking the relevant part, we have:
\begin{equation}
\begin{aligned}
&A_V(t) \mathcal{P} \chi(t) = \sum_\alpha s_\alpha(t) \rho_{\mathrm{S}}(t) \otimes B_\alpha(t) \tau_R \label{eq:APP}
\end{aligned}
\end{equation}
Taking the trace over both system and environment then leads to Eq.~\eqref{eq:rel_part}. For the part originating from $\mathcal{Q}\chi(t)$ we use Eq.~\eqref{eq:q_eom} to get:
\begin{equation}
\begin{aligned}
&A_V(t) \mathcal{Q} \chi(t) = \sum_\alpha s_\alpha(t) B_\alpha(t) \mathcal{Q} \chi(t)\\
& =  \sum_\alpha s_\alpha(t) \rho_{\mathrm{S}}(t_0) \otimes  B_\alpha(t)(\tau_B - \tau_R) \\
&- i  \lambda \sum_{i\alpha} \int_{0}^t \mathrm{~d} s \bigg \{  s_\alpha(t) A_i(s) \rho_\mathrm{S}(0) \otimes B_\alpha(t) \tau_i(s) \\
     & - s_\alpha(t) \rho_\mathrm{S}(0) A_i(s) \otimes \tau_\alpha(t) B_i(s) \bigg \}\\
&-i\lambda \sum_{\alpha i} \int_{0}^t \mathrm{~d} s \bigg \{ s_\alpha(t) A_i(s) \rho_\mathrm{S}(t) \otimes B_\alpha(t)B_i(s)\tau_R \\
&-s_\alpha(t) \rho_\mathrm{S}(t) A_i(s) \otimes B_\alpha(t)\tau_R B_i(s)\bigg \} \label{eq:APQ}
\end{aligned}
\end{equation}
where we defined $\tau_i(s) = B_i(s) \left ( \tau_B-\tau_R\right)$. Noting that $\trB{B_\alpha(t) \tau_i(s)} = C_{\alpha i}(t,s) - \expval{B_\alpha(t)} \Gamma_i(s)$, introducing the operators in Eqs.\eqref{eq:psi_alpha}-\eqref{eq:phi_alpha}, and taking the trace leads to Eq.~\eqref{eq:correction_terms}.

\subsection{Modified Quantum Regression theorem \label{app:QRT}}
As shown in Sec.~\ref{sec:me_for_two_time}, corrections to dynamics occur when considering the Quantum Regression theorem in the Projection Operator formalism. 

In the main text, we consider the linear response, which requires the quantity $S^{1}(t,\tau) = \mathrm{Im}\left( \trSB{{\mu^\dagger(t+\tau) \mu(t) \chi(0)}} \right)$. $S^{1}(t,\tau) = \trSB{\mu_P^\dagger \tilde{\Lambda}(t,\tau)}$ is a two-time expectation value and thus will follow the dynamics outlined in Sec.~\ref{sec:me_for_two_time}, which we here explicitly calculate.

We start by finding $\Lambda(t,0) = \mu_P (\mathcal{P} + \mathcal{Q}) \chi(t)$ with $\mu_P = \mu_P^\dagger = \sigma B_+ + \sigma^\dagger B_- = \sum_\alpha s_\alpha \otimes B_\alpha$. We notice that $\mathcal{P} \Lambda(t,0) = \mathcal{P}\left [\sum_\alpha s_\alpha(t) B_\alpha(t) \chi(t)\right]$. As in the above, we do not directly have access to $\chi(t)$, so we insert the identity $\mathcal{P} + \mathcal{Q}$ to obtain an expression with $\mathcal{P} \chi(t)$ and $\mathcal{Q}\chi(t)$. This leads exactly to Eqs.~\eqref{eq:APP} and \eqref{eq:APQ}. Applying $\mathcal{P}$, this gives:
\begin{equation}
\begin{aligned}
\mathcal{P} \tilde{\Lambda}(t,0) &= \bigg \{ \sum_\alpha \expval{B_\alpha} s_\alpha(t) \tilde{\rho}(t) + \sum _\alpha \Gamma_\alpha(t)  s_\alpha(t) \rho_\mathrm{S}(0)  \\ 
    & -i \sum_\alpha  s_\alpha(t) \Psi_\alpha^\mathcal{I}(t) \rho_\mathrm{S} (0)  - s_\alpha(t) \rho_\mathrm{S}(0) \Theta_\alpha^\mathcal{I}(t) \\
    &-i \sum_\alpha  s_\alpha(t) \Psi_\alpha(t) \tilde{\rho}_\mathrm{S} (t)  - s_\alpha(t) \tilde{\rho}_\mathrm{S}(t) \Theta_\alpha(t))\bigg \}  \otimes \tau_R  
\end{aligned}
\end{equation}
Now, we have not explicitly defined what the initial state $\chi(0)$ is. However, if $\mathcal{Q} \chi(0)=0$, then all the terms above related to $\rho_\mathrm{S}(0)$ are zero, which comes from the definition of $\Gamma_\alpha(t)$ and $C_{\alpha i}(t,s)$. If $\mathcal{Q} \chi(0) \neq 0$, but for example is the initial state considered in Eq.\eqref{eq:Qchi0}, then in the limit $t \rightarrow \infty$ it is easy to show that $\Gamma_\alpha(t) = 0$ and $C_{\alpha i }(t,s)=0$. Since we only consider the linear-response spectrum, which takes place in the limit of $t \rightarrow \infty$, we thus disregard the terms related to $\rho_\mathrm{S}(0)$. With this, we arrive at Eq.~\eqref{eq:P_lambda_t0}. The physical interpretation being that no matter the initial state (even of the bath), we relax to the equilibrium state in the linear response spectrum.

Similarly, we can apply $\mathcal{Q}$ to Eqs.~\eqref{eq:APP} and \eqref{eq:APQ} to get:
\begin{equation}
    \mathcal{Q} \Lambda(t,0) = \mathcal{Q} \mu_P(t) \mathcal{P} \chi(t) + \mathcal{Q}\mu_P(t)\mathcal{Q} \chi(t)
\end{equation}

which gives:

\begin{equation}
\begin{aligned}
    \mathcal{Q} \mu_P(t) \mathcal{P} \chi(t)=  \sum_\alpha s_\alpha \rho_\mathrm{S}(t) \otimes (B_\alpha(t) \tau_R-\expval{B_\alpha(t)}\tau_R), \label{eq:QPChi}    
\end{aligned}
\end{equation}

and

\begin{equation}
\begin{aligned}
&\mathcal{Q}\mu_P(t)\mathcal{Q}\chi(t)
=
\sum_\alpha
s_\alpha(t)\rho_\mathrm{S}(0)
\otimes
\bigg[
B_\alpha(t)(\tau_B-\tau_R)
\\
&\qquad\qquad\qquad
-
\trB{
B_\alpha(t)(\tau_B-\tau_R)
}
\tau_R
\bigg]
\\[0.5em]
&\quad
-i\lambda
\sum_{\alpha i}
\int_0^t \mathrm{d}s
\bigg\{
s_\alpha(t)A_i(s)\rho_\mathrm{S}(0)
\otimes
\bigg[
B_\alpha(t)B_i(s)(\tau_B-\tau_R)
\\
&\qquad\qquad\qquad\qquad
-
\trB{
B_\alpha(t)B_i(s)(\tau_B-\tau_R)
}
\tau_R
\bigg]
\\
&\qquad\qquad
-
s_\alpha(t)\rho_\mathrm{S}(0)A_i(s)
\otimes
\bigg[
B_\alpha(t)(\tau_B-\tau_R)B_i(s)
\\
&\qquad\qquad\qquad\qquad
-
\trB{
B_\alpha(t)(\tau_B-\tau_R)B_i(s)
}
\tau_R
\bigg]
\bigg\}
\\[0.5em]
&\quad
-i\lambda
\sum_{\alpha i}
\int_0^t \mathrm{d}s
\bigg\{
s_\alpha(t)A_i(s)\rho_\mathrm{S}(t)
\otimes
\bigg[
B_\alpha(t)B_i(s)\tau_R
\\
&\qquad\qquad\qquad\qquad
-
C_{\alpha i}(t,s)\tau_R
\bigg]
\\
&\qquad\qquad
-
s_\alpha(t)\rho_\mathrm{S}(t)A_i(s)
\otimes
\bigg[
B_\alpha(t)\tau_R B_i(s)
\\
&\qquad\qquad\qquad\qquad
-
C_{i\alpha}(s,t)\tau_R
\bigg]
\bigg\}
\\
&\quad
+
\mathcal{O}(\lambda^2).
\label{eq:QLambda_t0}
\end{aligned}
\end{equation}

Again, in the following, we can disregard terms relating to $\rho_\mathrm{S}(0)$ and $C_{\alpha,i}(t,s)$ using the same reasoning as above. We now need to apply $\mathcal{I}(\tau,0)$ to find the dynamics.   

For $\mathcal{I}(\tau,0) [\mathcal{Q} \mu_P(t) \mathcal{P} \chi(t)]$, we can simply use Eq.~\eqref{eq:me_inhomo}, with $\tau_B = B_\alpha(t) \tau_R$, which means that the correlation functions become $\Gamma_i(\tau) \rightarrow \Gamma_{i\alpha}(\tau,t)$:
\begin{equation}
\begin{aligned}
&\Gamma_{i\alpha}(\tau,t) =  \trB{B_i(\tau)(B_\alpha(t)\tau_R-\expval{B_\alpha(t)}\tau_R)} \\ 
&= C_{i\alpha}(\tau-t)- \expval{B_\alpha(t)}\expval{B_i(\tau)},  
\end{aligned}
\end{equation}
and $C_{ij}^{\mathcal{I}}(\tau,s)\rightarrow C_{ij\alpha}^{\mathcal{I}}(\tau,s,t) $
\begin{equation}
\begin{aligned}
C_{ij\alpha}^{\mathcal{I}}(\tau,s,t) 
&= \trB{B_i(\tau)B_j(s)\left(B_\alpha(t)\tau_R-\expval{B_\alpha(t)}\tau_R\right)}  \\
&= C_{ij\alpha}(\tau,s,t) - \expval{B_\alpha(t)} C_{ij}(\tau,s).
\end{aligned}
\end{equation}
Inserting these definitions in Eq.~\eqref{eq:me_inhomo}, we arrive at Eq.~\eqref{eq:I_QP}.

For $\mathcal{I}(\tau,0)[\mathcal{Q} \mu_P \mathcal{Q}\chi(t)]$, we only include contributions up until $\lambda^2$. This means that we only have the contribution from first order in $\lambda$ from the inhomogeneous propagator: $\mathcal{I}(\tau,0)[\mathcal{Q} \mu_P \mathcal{Q}\chi(t)] = -i \lambda[H_I(\tau), \mathcal{Q} \mu_P \mathcal{Q}\chi(t)]$. Inserting only the part without $\rho_\mathrm{S}(0)$ in Eq.~\eqref{eq:QLambda_t0} we get:

\begin{equation}
\begin{aligned}
    &\mathcal{I}(\tau,0)[\mathcal{Q} \mu_P \mathcal{Q}\chi(t)] = \\
     &- \lambda^2 \mathrm{tr}_\mathrm{B} \bigg [  H_I(\tau), \sum_{\alpha j} \int_{0}^t \mathrm{~d} s \bigg \{ s_\alpha(t) A_j(s) \rho_\mathrm{S}(t) \otimes B_\alpha(t)B_j(s)\tau_R \\
&-s_\alpha(t) \rho_\mathrm{S}(t) A_j(s) \otimes B_\alpha(t)\tau_R B_j(s) \bigg \} \bigg ] \\
& = -\lambda^2 \sum_{ij \alpha} \int_0^t d s \bigg  \{ \\
&\trB{B_i(\tau)B_\alpha(t)B_j(s)\tau_R} [A_i(\tau), s_\alpha(t)A_j(s) \rho_\mathrm{S}(t) ] \\
& - \trB{B_i(\tau)B_\alpha(t)\tau_R B_j(s)} \left [A_i(\tau), s_\alpha(t) \rho_\mathrm{S}(t) A_j(s) \right ] \bigg \}
\end{aligned}
\end{equation}
where we used that $\expval{B_i(t)} = 0$ for $i \in \{x,y,z\}$. Performing the trace over the bath then results in Eq.~\eqref{eq:I_QQ}

\subsection{Corrections to two-time expectation values\label{app:two_time}}
When considering the expectation value $S^1(t,\tau) = \trSB{\mu_P \tilde{\Lambda}(t,\tau)}$, we not only need to consider the changed dynamics of $\tilde{\Lambda}(t,\tau)$ introduced in Sec.~\ref{sec:me_for_two_time} and Appendix \ref{app:QRT}, but also corrections to the expectation value itself occur when $\mu_P$ is acting on both environment and system.  Inserting $\mathcal{Q} + \mathcal{P}$, we get:

\begin{equation}
    S^{1}(t,\tau)  = S^1_{rel}(t,\tau) + S^1_{irrel}(t,\tau),
\end{equation}
with
\begin{equation}
     S^1_{rel}(t,\tau)  = \trSB{\mu_P(\tau) \mathcal{P}\tilde{\Lambda}(t,\tau)}, 
\end{equation}
and 
\begin{equation}
     S^1_{irrel}(t,\tau) = \trSB{\mu_P(\tau) \mathcal{Q}\tilde{\Lambda}(t,\tau)}.
\end{equation}

Again, if we decompose $\mu_P(\tau) = \sum_\alpha s_\alpha(\tau) \otimes B_\alpha(\tau)$, we get:

\begin{equation}
     S^1_{rel}(t,\tau) = \sum_\alpha \expval{B_\alpha(\tau)} \trS{s_\alpha(\tau) \tilde{\rho}_{S}(t,\tau)}
\end{equation}
where $\tilde{\rho}_{S}(t,\tau) = \mathrm{e}^{iH_S \tau} \trB{\Lambda(t,\tau)} \mathrm{e}^{-iH_S \tau} $. 

The contribution from $S^1_{irrel}(t,\tau)$ can be split up into three parts, taking Eq.~\eqref{eq:QLambda} we have:
\begin{equation}
\begin{aligned} 
S^1_{irrel}(t,\tau)  &=  \trSB{\mu_P(\tau) \mathcal{Q} {\Lambda}(t,0)} \\
&+ \trSB{\mu_P(\tau) \int_{0}^{\tau} \mathrm{~d} s \mathcal{Q} \mathcal{L}\left(s\right) \mathcal{Q} {\Lambda}(t,0)} \\
&+\trSB{\mu_P(\tau) \int_{0}^{\tau} \mathrm{~d} s \mathcal{L}(s) \mathcal{P} \tilde{\Lambda}(t,\tau)}    
\end{aligned}
\end{equation}

The simplest is the last term, which is identical to the term we find for the one-time expectation values, thus:

\begin{equation}
\begin{aligned}
&\trSB{\mu_P(\tau) \int_{0}^{\tau} \mathrm{~d} s \mathcal{L}(s) \mathcal{P} \tilde{\Lambda}(t,\tau)}    = \\
    &-i \sum_\alpha \trS { s_\alpha(\tau) \Psi_\alpha(\tau) \tilde{\rho}_{S}(t,\tau)  - s_\alpha(\tau) \tilde{\rho}_{S}(t,\tau) \Theta_\alpha(\tau)) }    
\end{aligned}
\end{equation}

For the other two terms, we split the contribution from $\mathcal{Q}\Lambda(t,0)$ into its two parts: $\mathcal{Q}\Lambda(t,0) = \mathcal{Q} \mu_P(t)\mathcal{P}\chi(t) + \mathcal{Q} \mu_P(t)\mathcal{Q}\chi(t)$, which we defined in Eq.~\eqref{eq:QPChi} and \eqref{eq:QLambda_t0}. For $\mathcal{Q} \mu_P(t)\mathcal{P}\chi(t)$, we can simply re-use the terms relating to the inhomogeneous terms Eq.~\eqref{eq:correction_terms} but again with  $\Gamma_\alpha(\tau) \rightarrow \Gamma_{\alpha \beta}(\tau,t)$ and $C_{\alpha j}^{\mathcal{I}}(\tau,s)\rightarrow C_{\alpha j \beta}^{\mathcal{I}}(\tau,s,t) $
Inserting, we get:

\begin{equation}
\begin{aligned}
&\trSB{\mu_P(\tau) \mathcal{Q} \mu_P(t)\mathcal{P}\chi(t)} \\ 
&+ \trSB{\mu_P(\tau) \int_{0}^{\tau} \mathrm{~d} s \mathcal{Q} \mathcal{L}\left(s\right) \mathcal{Q} \mu_P(t)\mathcal{P}\chi(t)} \\
&= \sum _{\alpha \beta} \Gamma_{\alpha \beta}(\tau,t) \trS { s_\alpha(\tau) \rho_\mathrm{S}(t,0) } \\ 
    & -i \sum_{\alpha \beta} \trS { s_\alpha(\tau) \Psi_{\alpha \beta}^\mathcal{I}(\tau,t) \rho_\mathrm{S}(t,0)  - s_\alpha(\tau) \rho_\mathrm{S}(t,0) \Theta_{\alpha \beta}^\mathcal{I}(\tau,t)  }
\end{aligned}
\end{equation}
where now:
\begin{equation}
    \Psi_{\alpha \beta}^\mathcal{I}(\tau,t)  
    =  - \Phi_{\alpha \beta}(\tau,t) 
    + \sum_j \int_{0}^\tau \mathrm{~d} s \,
    C^\mathcal{I}_{\alpha j \beta }(\tau,s,t) A_j(s),  
    \label{eq:psi_I_alphabeta}
\end{equation}
\begin{equation}
    \Theta_{\alpha \beta}^\mathcal{I}(\tau,t)  
    = - \Phi_{\alpha \beta}(\tau,t) 
    +\sum_j \int_{0}^\tau \mathrm{~d} s \,
    C^\mathcal{I}_{j\alpha \beta}(s,\tau,t) A_j(s), 
    \label{eq:theta_I_alphabeta}
\end{equation}
\begin{equation}
    \Phi_{\alpha \beta}(\tau,t) 
    = \expval{B_\alpha(\tau)} 
    \sum_j \int_0^\tau ds \, \Gamma_{j \beta}(s,t) A_j(s). 
    \label{eq:phi_alphabeta}
\end{equation}

For the other part, we get:

\begin{equation}
\begin{aligned}
&\trSB{\mu_P(\tau) \mathcal{Q} \mu_P(t)\mathcal{Q}\chi(t)} \\
&= -i \sum_{\alpha \beta i} \int_0^t ds \bigg \{ C_{\alpha \beta i}(\tau,t,s) \trS{s_\alpha (\tau) s_\beta(t) A_i(s) \rho_\mathrm{S}(t) } \\ 
&- C_{ i \alpha \beta}(s,\tau,t) \trS{s_\alpha (\tau) s_\beta(t) \rho_\mathrm{S}(t) A_i(s)} \bigg \}, \label{eq:LR_ss_correction}   
\end{aligned}
\end{equation}

and finally,

\begin{equation}
\begin{aligned}
     &\trSB{\mu_P(\tau) \int_{0}^{\tau} \mathrm{~d} s \mathcal{Q} \mathcal{L}\left(s\right) \mathcal{Q} {\Lambda}(t,0)}  \\ 
     &= - \sum_{\alpha \beta i j} \int_0^{\tau} ds \int_0^t ds' \bigg \{ \\
     &C^\mathcal{I}_{\alpha j \beta i}(\tau,s,t,s') s_\alpha(\tau)A_j(s)s_\beta(t) A_i(s') \rho_\mathrm{S}(t) \\
     & - C^\mathcal{I}_{i\alpha j\beta}(s',\tau,s,t) s_\alpha(\tau)A_j(s)s_\beta(t)  \rho_\mathrm{S}(t)A_i(s') \\
     & -C^\mathcal{I}_{j \alpha  \beta i}(s,\tau,t,s') s_\alpha(\tau)s_\beta(t) A_i(s') \rho_\mathrm{S}(t)A_j(s)\\
     & +C^\mathcal{I}_{ij \alpha \beta}(s',s,\tau,t) s_\alpha(\tau) s_\beta(t)\rho_\mathrm{S}(t)  A_i(s')A_j(s) \bigg \} \label{eq:correction_omitted}
\end{aligned}
\end{equation}
where we defined $C^\mathcal{I}_{\alpha \beta i j} = C_{\alpha j \beta i}(\tau,s,t,s') - \expval{B_\alpha(\tau)} C_{j \beta i}(s,t,s')$ and $C_{\alpha j \beta i}(\tau,s,t,s')  = \trB{B_\alpha(\tau)B_j(s)B_\beta(t)B_i(s') \tau_R}$. We note that this requires a fourth operator expectation value with respect to the environment. With the method showcased in Appendix \ref{app:correlation_functions}, this is, in principle, possible to calculate, albeit tedious. We found from running the simulations that the correction in Eq.~\eqref{eq:LR_ss_correction} was very short-lived and small. The correction in Eq.~\eqref{eq:correction_omitted} would similarly be relevant on short time scales; additionally, it is of second order in $\Lambda$, and thus would be even smaller in magnitude. For these reasons, we have chosen to omit this correction in the calculations presented in the paper.

Thus, in total, we arrive at:

\begin{equation}
\begin{aligned} 
&S^1_{irrel}(t,\tau)  = \\
&-i \sum_\alpha \trS { s_\alpha(\tau) \Psi_\alpha(\tau) \tilde{\rho}_{S}(t,\tau)  - s_\alpha(\tau) \tilde{\rho}_{S}(t,\tau) \Theta_\alpha(\tau)) } \\
&+\sum _{\alpha \beta} \Gamma_{\alpha \beta}(\tau,t) \trS { s_\alpha(\tau) \rho_\mathrm{S}(t,0) } \\ 
    & -i \sum_{\alpha \beta} \trS { s_\alpha(\tau) \Psi_{\alpha \beta}^\mathcal{I}(\tau,t)  \rho_\mathrm{S}(t,0)  - s_\alpha(\tau) \rho_\mathrm{S}(t,0) \Theta_{\alpha \beta}^\mathcal{I}(\tau,t)  }\\
    &-i \sum_{\alpha \beta i} \int_0^t ds \bigg \{ C_{\alpha \beta i}(\tau,t,s) \trS{s_\alpha (\tau) s_\beta(t) A_i(s) \rho_\mathrm{S}(t) } \\ 
&- C_{ i \alpha \beta}(s,\tau,t) \trS{s_\alpha (\tau) s_\beta(t) \rho_\mathrm{S}(t) A_i(s)} \bigg \}.
\end{aligned}
\end{equation}

\section{Calculating correlation functions of the environment}
\label{app:env_corr}

In this section, we calculate correlation functions of the bosonic environment required throughout the paper. We introduce a few tricks that are useful when dealing with higher-order correlations involving more than two operators. 

In this paper, we consider the unbiased spin-boson model with the system coupling operator $\sigma_z$ in Eq.~\eqref{eq:H}, if instead one considers a coupling of the form $\sigma^\dagger\sigma$, as is common in models of quantum dots \cite{Nazir2016ModellingDots,Iles-Smith2017PhononSources,Iles-Smith2017LimitsEmitters,Denning2020PhononSources,Bundgaard-Nielsen2021Non-MarkovianElectrodynamics}, this slightly alters the Hamiltonian in Eq.~\eqref{eq:HV} with the introduction of a polaron shift, and also the magnitude of the correlation functions which we are about to derive will differ by factors of $1$, $2$, and $4$ depending on how many $B_\pm$ operators are involved, see Ref.~\citenum{Nazir2016ModellingDots} for details. 

We begin by recalling a few useful properties of the bosonic displacement operator
\begin{equation}
 \ket{\alpha} = D(\alpha) \ket{0}, \ \mathrm{with} \ \ D(\alpha) = \exp\!\left(\alpha b^\dagger - \alpha^* b \right),
\end{equation}
where $b$ and $b^\dagger$ are the bosonic annihilation and creation operators.

The product of two displacement operators satisfies \cite{Gerry2004IntroductoryOptics}
\begin{equation}
D(\alpha) D(\beta) = \exp \left(i \operatorname{Im}\left(\alpha \beta^*\right)\right)D(\alpha + \beta),
\label{eq:Dprod}
\end{equation}

Finally, a key identity is the transformation of the annihilation operator under conjugation by a displacement operator \cite{Nazir2016ModellingDots},
\begin{equation}
D^\dagger(\alpha) \, b \, D(\alpha) = b + \alpha,
\qquad
D^\dagger(\alpha) \, b^\dagger \, D(\alpha) = b^\dagger + \alpha^*.
\label{eq:Dshift}
\end{equation}

\subsection{Two-time correlation functions}
We start by considering the correlation function involved in the standard variational ME. 

\subsection{$C_{ZZ}$}

We here need to calculate $C_{ZZ}(t-s) = \expval{B_Z(t)B_Z(s)}$, considering only the single bosonic mode, we can later generalize to the many-mode case. Thus $\tilde{B}_Z(t) = (g_\mathbf{k}-f_\mathbf{k})(b_\mathbf{k} \mathrm{e}^{-i \nu_\mathbf{k}t}+ b_\mathbf{k}^\dagger \mathrm{e}^{i \nu_\mathbf{k}t})$ and we have:
\begin{equation}
\begin{aligned}
    &\expval{\tilde{B}_Z(t)\tilde{B}_Z(s)} \\ 
    &= \left (g_\mathbf{k}-f_\mathbf{k}\right)^2 \expval{(b_\mathbf{k} \mathrm{e}^{-i \nu_\mathbf{k}t}+ b_\mathbf{k}^\dagger \mathrm{e}^{i \nu_\mathbf{k}t})(b_\mathbf{k} \mathrm{e}^{-i \nu_\mathbf{k}s}+ b_\mathbf{k}^\dagger \mathrm{e}^{i \nu_\mathbf{k}s})} \\
    & = \left ( g_\mathbf{k}-f_\mathbf{k} \right)^2 \left ( \Bar{n}_\mathbf{k} \mathrm{e}^{i \nu_\mathbf{k}(t-s)}  +(\Bar{n}_\mathbf{k}+1) \mathrm{e}^{- i \nu_\mathbf{k}(t-s)} \right)
\end{aligned}
\end{equation}
where we used that the environment is in the thermal state and thus $\expval{b_\mathbf{k}^\dagger b_\mathbf{k}} = \Bar{n}_\mathbf{k}$, $\expval{b_\mathbf{k} b_\mathbf{k}^\dagger} = \Bar{n}_\mathbf{k} +1$, $\expval{b_\mathbf{k} b_\mathbf{k}} = 0$, and $\expval{b_\mathbf{k}^\dagger b_\mathbf{k}^\dagger} = 0$. Using that $n_\mathbf{k} =  \left ( {e^{\beta \hbar \nu_\mathbf{k}} - 1} \right)^{-1}$
and $2 n_\mathbf{k} + 1 = \coth\!\left(\frac{\beta \hbar \nu_\mathbf{k}}{2}\right)$, we arrive in the many-mode limit:
\begin{equation}
\begin{aligned}
    C_{ZZ}(t-s) = &\int_0^\infty d \nu J(\nu) (1-F(\nu))^2   \\
    &\big[ \cos(\nu (t-s)) \coth(\beta \nu/2) - i \sin(\nu (t-s)) \big ]    
\end{aligned}
\end{equation}

\subsection{$C_{XX}$ and $C_{YY}$}

Now when we want to compute $C_{XX}(t-s)$ and $C_{YY}(t-s)$, we are interested in correlations of the form $\expval{B_\pm(t) B_\pm(s)}$ and $\expval{B_\pm(t) B_\mp(s)}$.

We start by considering the single mode case and write: ${B}_+(t) = D(h(t)) = \exp[x(t)]$, with $h(t) =2 f_\mathbf{k}/\nu_\mathbf{k} e^{i\nu_\mathbf{k} t}$, and $x(t) = 2 f_\mathbf{k}/\nu_\mathbf{k} \left(b_\mathbf{k}^\dagger e^{i\nu_\mathbf{k} t} - b_\mathbf{k} e^{-i\nu_\mathbf{k} t}\right)$. From Ref.~\citenum{Nazir2016ModellingDots}, we already know the expectation value of a displacement operator with a thermal state

\begin{equation}
\langle D(h)\rangle
=\exp\!\left[-\frac{|h|^2}{2}\coth\!\left(\frac{\beta\omega}{2}\right)\right].
\label{eq:D_thermal}
\end{equation}

The generalization of the above to the many-mode case leads to the Franck-Condon factor: 

\begin{equation}
B=\langle B_\pm\rangle
=\exp\!\left[-2 \int_0 ^\infty d \nu \frac{F(\nu)^2 J(\nu)}{\nu^2}\coth\!\left(\frac{\beta\nu}{2}\right)\right],
\end{equation}

Similarly, we find $B_\pm(t) B_\pm(s) = \exp(f_\mathbf{k}/\nu_\mathbf{k} \sin(\nu_\mathbf{k} (t-s)))D(h(t)+h(s))$. Using the fact that $\abs{ h(t) + h(s)}^2 = 8 f_\mathbf{k}^2/\nu_\mathbf{k}^2 (1 + \cos(\nu_\mathbf{k} (t-s))$ we arrive at the many-mode generalization: $\expval{{B}_\pm(t) {B}_\pm(s)} = \expval{B}^2 \exp{-\phi(t-s)}$ and $\expval{{B}_\pm(t) {B}_\mp(s)} = \expval{B}^2 \exp{\phi(t-s)}$ with

\begin{equation}
\begin{aligned}
\phi(\tau) =&\int\limits_0^\infty\dd{\nu} \frac{4 J(\nu)F^2(\nu)}{\nu^2}[\coth(\frac{\beta\hbar\nu}{2})\cos(\nu\tau) -i\sin(\nu\tau)]\\
\end{aligned}
\end{equation}

With the definitions of $B_X$ and $B_Y$, we similarly get:

\begin{align}
C_{XX}(\tau) &= \frac{\ev{B}^2}{2}(e^{\phi(\tau)} + e^{-\phi(\tau)} - 2), \\
C_{YY}(\tau) &= \frac{\ev{B}^2}{2}(e^{\phi(\tau)} - e^{-\phi(\tau)}), \\
\end{align}

\subsection{$C_{YZ}$}

To calculate $C_{YZ}$, we need $C_{+Z}(t-s) = \langle B_+(t) B_Z(s) \rangle$. Again, we first consider a single bosonic mode and suppress the mode index. Writing
${B}_+(t) = \exp(x(t))$, with $x(t) =2 f_\mathbf{k}/\nu_\mathbf{k} \left(b_\mathbf{k}^\dagger e^{i\nu_\mathbf{k} t} - b_\mathbf{k} e^{-i\nu_\mathbf{k} t}\right)$. We expand the displacement operator as
\begin{equation}
\tilde{B}_+(t) = 1 + x(t) + \frac{1}{2}x^2(t) + \cdots .
\end{equation}

Since $\tilde{B}_Z$ is linear in the bosonic operators and the thermal state is Gaussian, we can use Wick's theorem to calculate the many-operator expectation values, where all odd-numbered expectation values have no contributions. For four linear operators, this amounts to $\expval{ABCD}  = \expval{AB}\expval{CD} + \expval{AC}\expval{BD}+ \expval{AD}\expval{BC}$. Thus, we get:

\begin{equation}
\begin{aligned}
    &\expval{\tilde{B}_+(t)\tilde{B}_z(s) } \\
    &= \expval{x(t) \tilde{B}_Z(s)} + \frac{1}{3!}\expval{x(t)^3 \tilde{B}_Z(s)}  + \cdots \\
    &= \expval{x(t) \tilde{B}_Z(s)}\left( 1 + \frac{1}{2!}(\expval{x(t)^2}) + \cdots  \right)
\end{aligned}
\end{equation}
Now, we recognize the part inside the parentheses as $\expval{\exp(x(t))}  = 1 + \expval{x(t)^2}/2! + \cdots$, i.e., the Franck-Condon factor (when taken in the many-mode limit). Similarly, we have:
\begin{equation}
\begin{aligned}
    &\langle x(t) \tilde{B}_Z(s) \rangle
     \\
    & =   2 \frac{f_\mathbf{k}(g_\mathbf{k}-f_\mathbf{k})}{\nu_\mathbf{k}}
    \left\langle 
    (b_\mathbf{k}^\dagger \mathrm{e}^{i \nu_\mathbf{k}t} - b_\mathbf{k} \mathrm{e}^{-i \nu_\mathbf{k}t})(b_\mathbf{k} \mathrm{e}^{-i \nu_\mathbf{k}s}+ b_\mathbf{k}^\dagger \mathrm{e}^{i \nu_\mathbf{k}s})
    \right\rangle \nonumber \\
    &= 2 \frac{f_\mathbf{k}(g_\mathbf{k}-f_\mathbf{k})}{\nu_\mathbf{k}}\left ( n_\mathbf{k} \mathrm{e}^{i \nu_\mathbf{k}(t-s)} - ( n_\mathbf{k} +1)\mathrm{e}^{-i \nu_\mathbf{k}(t-s)}   \right) \\
    & =  2 \frac{f_\mathbf{k}(g_\mathbf{k}-f_\mathbf{k})}{\nu_\mathbf{k}}\left ( i \coth(\beta \nu_\mathbf{k}/2) \sin(\nu_\mathbf{k}(t-s)) - \cos(\nu_\mathbf{k}(t-s))  \right)
\end{aligned}
\end{equation}

By combining the above with the many-mode limit, we arrive at:
\begin{equation}
\begin{aligned}
    C_{+Z}(t-s) &= 2 \ev{B}\int_0^\infty \dd{\nu} J(\nu)\nu^{-1}F(\nu)[1-F(\nu)] \cdot \\ 
    &[ i \coth(\beta\nu/2)\sin(\nu(t-s)) - \cos(\nu(t-s))], \\
\end{aligned}
\end{equation}

Similarly,

\begin{align}
    C_{Z+}(t-s) &= -C_{+Z}(t-s)  \\
    C_{-Z}(t-s) &= -C_{+Z}(t-s) \\
    C_{Z-}(t-s) &= C_{+Z}(t-s)
\end{align}

Now using the definition of $B_Y$, we finally get:
\begin{equation}
\begin{aligned}
    C_{YZ}(t-s) &= i C_{+Z}(t-s) \\ 
    &=  -2 \ev{B}\int_0^\infty \dd{\nu} J(\nu)\nu^{-1}F(\nu)[1-F(\nu)] \cdot \\ 
    &[ \coth(\beta\nu/2)\sin(\nu(t-s)) + i\cos(\nu(t-s))], \\
\end{aligned}
\end{equation}

and $C_{ZY}(t-s) = - C_{YZ}(t-s)$.

We note that the remaining correlation functions $C_{XY},\;C_{YX},\;C_{XZ}$ and $C_{ZX}$ are zero, by construction.

\subsection{Inhomogeneous corrections  \label{app:correlation_functions}}

We now wish to calculate the inhomogeneous terms introduced in Sec.~\ref{sec:me}. We assume that the initial state of the bath in the untransformed frame is simply $\tau_B = \tau_R$, because the variational master equation is taking place in the transformed frame, which leads to $\mathcal{Q} \chi(t_0) \neq 0$.

To calculate the new correlation functions, we assume that we have the TLS initially in $\rho_S(0) = \dyad{1}$ and utilize that:
\begin{equation}
\begin{aligned}
\mathcal{Q} {\chi}(t_0)& = (I - \mathcal{P}) ( \tilde B_- \ket{1}\bra{1} \otimes  \tau_B  \tilde  B_+) \\
&= \ket{1}\bra{1} \otimes ( \tilde B_- \tau_R \tilde B_+ - \tau_R)    
\end{aligned}
\end{equation}

It is helpful to realize:
\begin{equation}
    \mathrm{tr}_B \left(  B_i  \tilde B_- \tau_R  \tilde B_+ \right ) = \mathrm{tr}_B \left(  B_{IV} \tau_{B} \right ) 
\end{equation}
where we defined $B^\mathcal{I}_{i} =  \tilde B_+ B_i  \tilde B_-$. Similarly:
\begin{equation}
    \mathrm{tr}_B \left(  B_i(\tau)B_j(0) \tau_{BV} \right ) = \mathrm{tr}_B \left(  B^\mathcal{I}_{i}(\tau)B^\mathcal{I}_{j}(0) \tau_{B} \right )
\end{equation}
Following \cite{McCutcheon2011ConsistentEquation}, we have:
\begin{equation}
    B_{\pm}^\mathcal{I} = B_\pm \mathrm{e}^{\pm i \psi(t)} 
\end{equation}
where 
\begin{equation}
    \psi(t)= - 4\int_0^\infty d \omega \frac{J(\omega)}{\omega^2}F(\omega)^2 \sin(\omega t) = \mathrm{Im}(\phi(t))
\end{equation}
where it follows then that
\begin{equation}
    B^\mathcal{I}_{X}(t) = \frac{1}{2}( B_+(t) \mathrm{e}^{i \psi(t)}  + B_-(t) \mathrm{e}^{- i \psi(t)} - 2 \expval{B}) \label{eq:b_xv}
\end{equation}
\begin{equation}
    B^\mathcal{I}_{Y}(t) = \frac{i}{2}( B_+(t) \mathrm{e}^{i \psi(t)}  - B_-(t) \mathrm{e}^{- i \psi(t)})\label{eq:b_yv}
\end{equation}
Similarly,
\begin{equation}
    B^\mathcal{I}_{Z}(t)  = B_Z(t) + Z(t) \label{eq:b_zv}
\end{equation}
where
\begin{equation}
\begin{aligned}
    Z(t) &= - 2 \int_0^\infty d \omega \frac{J(\omega)}{\omega}F(\omega)(1-F(\omega))\cos(\omega t) \\
    &= \mathrm{Im}(C_{YZ}(t))/\expval{B}    
\end{aligned}
\end{equation}
With this, we get:
\begin{align}
\Gamma_\pm(t) &= \expval{B} (\mathrm{e}^{\pm i  \psi(t)}-1) \\
\Gamma_X(t) &= \expval{B} (\cos(\psi(t)) - 1) \\
\Gamma_Y(t) &= - \expval{B} \sin(\psi(t)) \\
\Gamma_Z(t) &= Z(t)
\end{align}
\begin{equation}
\begin{aligned}
C_{XX}^{\mathcal{I}}(\tau) = &\frac{\expval{B}^2}{2} \bigg [ \mathrm{e}^{\phi(\tau)} \left ( \cos[\psi(\tau) - \psi(0)] - 1 \right) \\
&+ \mathrm{e}^{-\phi(\tau)} \left ( \cos[\psi(\tau) + \psi(0)] - 1 \right) \\
&-2 (\cos(\psi(\tau))+\cos(\psi(0))-2) \bigg ]  
\end{aligned}
\end{equation}
 \\
\begin{equation}
\begin{aligned}
C_{YY}^{\mathcal{I}}(\tau) &= \frac{\expval{B}^2}{2} \bigg[ 
e^{\phi(\tau)} \left( \cos[\psi(\tau) - \psi(0)] - 1 \right) 
\\
&- e^{-\phi(\tau)} \left( \cos[\psi(\tau) + \psi(0)] - 1 \right) \bigg]
\end{aligned}
\end{equation}

\begin{equation}
\begin{aligned}
C_{ZZ}^{\mathcal{I}}(\tau) =Z(\tau)Z(0)
\end{aligned}
\end{equation}

\begin{equation}
\begin{aligned}
C_{XY}^{\mathcal{I}}(\tau) &= \frac{\expval{B}^2}{2} \bigg[ 
e^{\phi(\tau)} \sin[\psi(\tau) - \psi(0)] 
\\&- e^{-\phi(\tau)} \sin[\psi(\tau) + \psi(0)] 
+ 2 \sin (\psi(0)) \bigg]
\end{aligned}
\end{equation}
\begin{equation}
\begin{aligned}
C_{YX}^{\mathcal{I}}(\tau) &= \frac{\expval{B}^2}{2} \bigg[ 
- e^{\phi(\tau)} \sin[\psi(\tau) - \psi(0)] \\  
& - e^{-\phi(\tau)} \sin[\psi(\tau) + \psi(0)] 
+ 2 \sin (\psi(\tau))  \bigg]
\end{aligned}
\end{equation}
\begin{equation}
\begin{aligned}
C_{ZX}^{\mathcal{I}}(\tau) &= -C_{YZ}(\tau) \sin (\psi(0)) + \Gamma_Z(\tau) \Gamma_X(0)
\end{aligned}
\end{equation}

\begin{equation}
\begin{aligned}
C_{XZ}^{\mathcal{I}}(\tau) &= C_{YZ}(\tau) \sin( \psi(\tau) )+\Gamma_Z(0) \Gamma_X(\tau)
\end{aligned}
\end{equation}

\begin{equation}
\begin{aligned}
C_{ZY}^{\mathcal{I}}(\tau) &= -C_{YZ}(\tau) [\cos (\psi(0)) - 1] + \Gamma_Z(\tau) \Gamma_Y(0)
\end{aligned}
\end{equation}

\begin{equation}
\begin{aligned}
C_{YZ}^{\mathcal{I}}(\tau) &= C_{YZ}(\tau) [\cos (\psi(\tau)) - 1] + \Gamma_Z(0) \Gamma_Y(\tau)
\end{aligned}
\end{equation}

\subsection{Higher-order correlation functions}

For the calculations involving the quantum regression theorem, we need correlation functions of the type $C_{i \alpha j}(t,s,\tau)$,$C_{i j \alpha}(t,s,\tau)$, $C_{i \alpha \beta}(t,s,\tau)$, $C_{ \alpha \beta i}(t,s,\tau)$, $C_{ \alpha i \beta}(t,s,\tau)$, and $C_{ i \alpha \alpha}(t,s,\tau)$ with $i,j \in \{X,Y,Z\}$ and $\alpha, \beta \in \{+,-\}$. We start by considering correlation functions of the type $C_{\pm \pm \pm}(s,\tau,t)$. Correlation functions involving just displacement operators can be computed by combining them using Eq.\eqref{eq:Dprod}. Doing so, one finds:

\begin{equation}
\begin{aligned}
    &C_{\pm \pm \pm}(t,s,\tau)
    = \expval{B_\pm(t)  B_\pm(s) B_\pm(\tau)}  \\
    &= \expval{B}^3 
       \mathrm{e}^{ - \phi(t- s)} 
       \mathrm{e}^{ - \phi(t- \tau)} 
       \mathrm{e}^{ - \phi(s- \tau)}
\end{aligned}    
\end{equation}

Similarly,
\begin{align}
     &C_{\pm \pm \mp}(t,s,\tau)  
     = \expval{B}^3 
       \mathrm{e}^{ - \phi(t- s)} 
       \mathrm{e}^{ \phi(t- \tau)} 
       \mathrm{e}^{ \phi(s- \tau)} \\
     &C_{\pm \mp \mp}(t,s,\tau)  
     = \expval{B}^3 
       \mathrm{e}^{ \phi(t- s)} 
       \mathrm{e}^{ \phi(t- \tau)} 
       \mathrm{e}^{- \phi(s- \tau)} \\
     &C_{\mp \pm \pm}(t,s,\tau)  
     = \expval{B}^3 
       \mathrm{e}^{ \phi(t- s)} 
       \mathrm{e}^{ \phi(t- \tau)} 
       \mathrm{e}^{- \phi(s- \tau)} \\
     &C_{\mp \mp \pm}(t,s,\tau)  
     = \expval{B}^3 
       \mathrm{e}^{-\phi(t- s)} 
       \mathrm{e}^{ \phi(t- \tau)} 
       \mathrm{e}^{\phi(s- \tau)} \\
     &C_{\pm \mp \pm}(t,s,\tau)  
     = \expval{B}^3 
       \mathrm{e}^{\phi(t- s)} 
       \mathrm{e}^{- \phi(t- \tau)} 
       \mathrm{e}^{\phi(s- \tau)}
\end{align}

Together with the definitions of $B_X$ and $B_Y$, we can get all combinations of $C_{i \alpha j}(\tau,t,s)$,$C_{i j \alpha}(t,s,\tau)$, $C_{i \alpha \beta}$, $C_{ \alpha \beta i}(t,s,\tau)$, $C_{ \alpha i \beta}(t,s,\tau)$, and $C_{ i \alpha \alpha}(t,s,\tau)$ with $i,j \in \{X,Y\}$ and $\alpha,\beta \in \{+,-\}$.

Similarly, we need correlations of the type $C_{Z\alpha\beta}$, $C_{\alpha\beta Z}$, and $C_{\alpha Z\beta}$, where $\alpha,\beta\in\{+,-\}$. It is useful to identify the labels $\alpha,\beta$ with the signs $\alpha,\beta=\pm1$, such that $B_\alpha=B_+$ for $\alpha=+1$ and $B_\alpha=B_-$ for $\alpha=-1$. Combining the displacement operators using Eq.~\eqref{eq:Dprod}, one obtains the compact relations
\begin{align}
C_{Z\alpha\beta}(t,s,\tau)
&=
i\expval{B}
\mathrm{e}^{-\alpha\beta\phi(s-\tau)}
\notag\\
&\quad\times
\left[
\alpha C_{YZ}(t-s)
+
\beta C_{YZ}(t-\tau)
\right],
\\
C_{\alpha\beta Z}(t,s,\tau)
&=
-i\expval{B}
\mathrm{e}^{-\alpha\beta\phi(t-s)}
\notag\\
&\quad\times
\left[
\alpha C_{YZ}(t-\tau)
+
\beta C_{YZ}(s-\tau)
\right],
\\
C_{\alpha Z\beta}(t,s,\tau)
&=
i\expval{B}
\mathrm{e}^{-\alpha\beta\phi(t-\tau)}
\notag\\
&\quad\times
\left[
-\alpha C_{YZ}(t-s)
+
\beta C_{YZ}(s-\tau)
\right].
\end{align}

With this all combinations of $C_{Z \alpha j}(\tau,t,s)$,$C_{i \alpha Z}(\tau,t,s)$, $C_{Z j \alpha}(\tau,t,s)$, and $C_{i Z \alpha}(\tau,t,s)$ with $i,j \in \{X,Y\}$ and the definitions of $B_X$ and $B_Y$, can be constructed, as well with $C_{Z \alpha \beta}(\tau,t,s)$, $C_{ \alpha \beta Z}(\tau,t,s)$, $C_{ \alpha Z \beta}(\tau,t,s)$, and $C_{ i \alpha \alpha}(\tau,t,s)$.

Finally, we need $C_{Z \alpha  Z}(\tau,t,s)$ and $C_{Z  Z \alpha}(\tau,t,s)$. For this, we need to use Wick's theorem together with the expansion of the displacement operator. As before, we start by considering the single-mode case:

\begin{equation}
\begin{aligned}
    &\expval{\tilde B_Z(\tau) \tilde B_Z(t) B_+(s)} = \expval{\tilde B_Z(\tau) \tilde B_Z(t)} \\ 
    &+ \frac{\expval{\tilde B_Z(\tau) \tilde B_Z(t) x(s)^2}}{2!} + \frac{\expval{\tilde B_Z(\tau) \tilde B_Z(t) x(s)^4}}{4!} + \cdots =  \\
    & \expval{\tilde B_Z(\tau) \tilde B_Z(t)} +  \frac{\expval{\tilde B_Z(\tau) \tilde B_Z(t)} \expval{x(s)^2}}{2!} \\ 
    &+ 2 \frac{\expval{\tilde B_Z(\tau) x(s)} \expval{ \tilde B_Z(t) x(s)}}{2!} + \frac{\expval{\tilde B_Z(\tau) \tilde B_Z(t)} \expval{x(s)^4}}{4!}  \\
    &+  12 \frac{\expval{\tilde B_Z(\tau) x(s)} \expval{ \tilde B_Z(t) x(s)}\expval{x(s)^2}}{4!} + \cdots =  \\
    &\expval{\tilde B_Z(\tau) \tilde B_Z(t)} \left (1 + \frac{\expval{x(s)^2}}{2!} + \frac{\expval{x(s)^4}}{4!} + \cdots \right )  \\ 
    &+ \expval{\tilde B_Z(\tau) x(s)} \expval{ \tilde B_Z(t) x(s)} \left ( 1 +  \frac{\expval{x(s)^2}}{2}  + \cdots \right ) 
\end{aligned}
\end{equation}

We notice that in the many-mode limit then the infinite sum of $x(t)$ is exactly the Franck-Condon factor $\expval{B} = \prod_\mathbf{k} \left (1 + \frac{\expval{x_k^2}}{2!} + \frac{\expval{x_k^4}}{4!} + \cdots \right )$, where $x_k = 2 f_\mathbf{k}/\nu_\mathbf{k} \left(b_\mathbf{k}^\dagger  - b_\mathbf{k} \right)$. With this, we get:

\begin{equation}
    C_{ZZ+}(t,s,\tau) = \expval{B} C_{ZZ}(t-s) - C_{YZ}(t,\tau) C_{YZ}(s,\tau) / \expval{B} ,
\end{equation}

Similarly, we get:

\begin{align}
    &C_{ZZ-}(t,s,\tau) = \expval{B} C_{ZZ}(t-s) - C_{YZ}(t,\tau) C_{YZ}(s,\tau) / \expval{B}, \\
    &C_{Z+Z}(t,s,\tau) = \expval{B} C_{ZZ}(t-\tau) + C_{YZ}(t,s) C_{YZ}(s,\tau) / \expval{B}, \\
    &C_{Z-Z}(t,s,\tau) = \expval{B} C_{ZZ}(t-\tau) + C_{YZ}(t,s) C_{YZ}(s,\tau) / \expval{B}
\end{align}

\section{Numerical details on Tensor-Network calculations\label{app:tensor_details}}

For the tensor-network calculations in Sections ~\ref{sec:populations} and \ref{sec:linear_response}, convergence of the algorithm relies on carefully tuning three primary numerical parameters to achieve the desired precision. First, the environmental memory time, $T_\mathrm{mem}$, must be sufficiently long to ensure that the bare environmental correlation function ($C_{ZZ}(\tau)$ with $f_\mathbf{k} = 0$) has fully decayed. Because the approach in Ref.~\citenum{Link2024OpenContraction} utilizes infinite uniform matrix product states, the computational overhead of extending $T_\mathrm{mem}$ beyond this relaxation time is minimal. Consequently, we conservatively set $T_\mathrm{mem} = \{1000/\Delta, 1000/\Delta, 1000/\Delta, 200/\Delta\}$ for the four studied spectral densities with Ohmicities $s=1,~1.5,~2,$ and 3, respectively. Second, the integration time step, $\Delta t$, must be small enough to adequately resolve the system dynamics and suppress errors originating from the second-order Trotter decomposition. We achieve convergence with $\Delta t = 0.02/\Delta$; further decreasing the step size to $\Delta t = 0.01/\Delta$ yields no observable changes in the dynamics. Finally, the truncation threshold, $\epsilon$, dictates the singular-value cutoff used to compress the matrix product state representing the environment. It must be chosen small enough to ensure the reduced system dynamics are fully converged. We find that a threshold of $\epsilon = 5 \times 10^{-10}$ is sufficient when using $\Delta t = 0.02/\Delta$. Rigorous testing with a tighter threshold of $\epsilon = 10^{-10}$ at $\Delta t = 0.01/\Delta$ confirmed that no significant numerical artifacts remain.

\section{Numerical cost of the Master Equations \label{app:numerical_costs}}
In this appendix, we provide details on the numerical costs of the master equation simulations. Specifically, we address the added complexity that the inhomogeneous source terms introduce. In table \ref{tab:numerical_costs} we provide the simulation times for the different master equation approaches with and without the inhomogeneous source terms for the case of $s=3$, where we have the largest coherence time and therefore also have the largest simulation time.

\begin{table*}[!htp]
    \centering
    \caption{Timing totals for the different master equation approaches with and without inhomogeneous source terms on the $s=3$ spectral density. The timings are performed on an Intel(R) Xeon(R) Gold 6126 CPU 2.60GHz processor. }
    \label{tab:numerical_costs}
    \begin{tabular}{lcccc}
        \toprule
        Method & Linear response & Coherences & Correction & Total corr. eval. \\
        \midrule
        Variational w/o inhomo.
            & $7.791 \,\mathrm{s}$ 
            & $2.036\,\mathrm{s}$ 
            & $0.036\,\mathrm{s}$ 
            & $2.072\,\mathrm{s}$ \\

        Variational w. inhomo.
            & $131.845\,\mathrm{s}$ 
            & $64.646\,\mathrm{s}$ 
            & $3.667\,\mathrm{s}$ 
            & $68.313\,\mathrm{s}$ \\

        Weak-coupling w/o inhomo.
            & $7.091\,\mathrm{s}$ 
            & $2.272\,\mathrm{s}$ 
            & -- 
            & $2.272\,\mathrm{s}$ \\

        Weak-coupling w. inhomo.
            & $101.694\,\mathrm{s}$ 
            & $2.272\,\mathrm{s}$ 
            & -- 
            & $2.272\,\mathrm{s}$ \\
        \bottomrule
    \end{tabular}
\end{table*}

\section{$s=1$ with weaker bath coupling $\alpha$\label{app:s1_weak}}
\begin{figure}
    \centering
    \includegraphics[width=\linewidth]{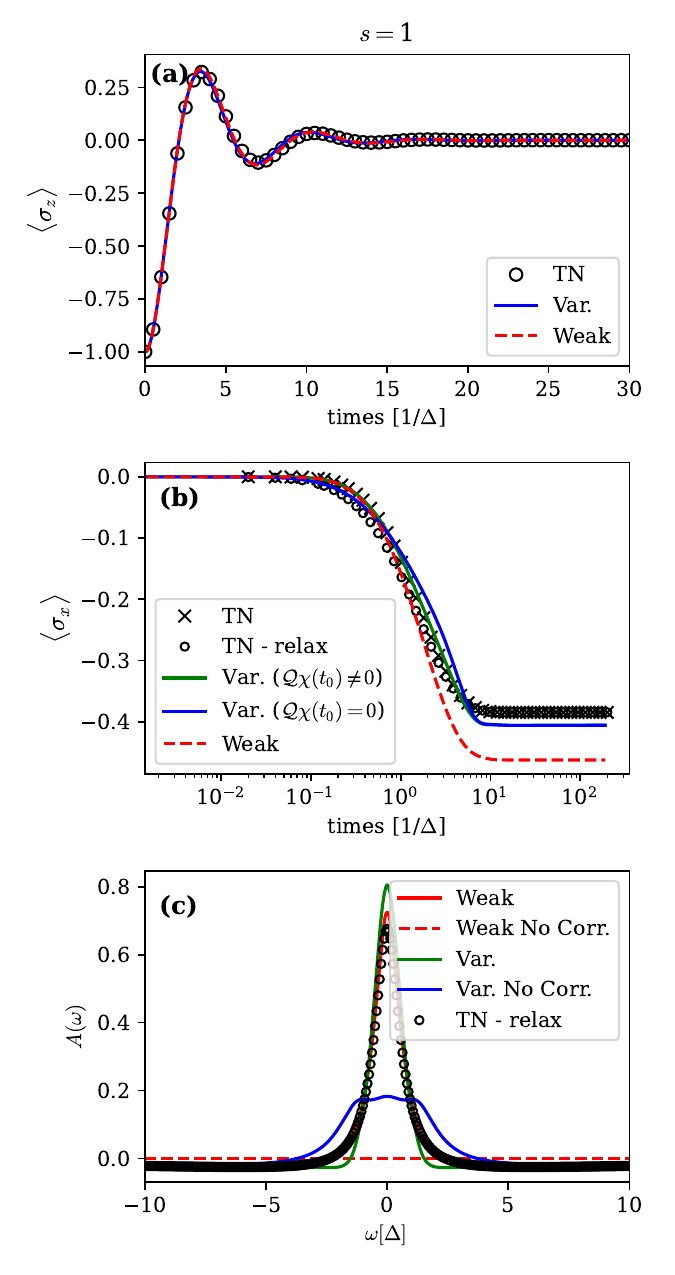}
    \caption{(a) Same as in Fig.~\ref{fig:excited_false} with $s=1$ but with $\alpha=0.05$, (b), same as in Fig.~\ref{fig:ground} with $s=1$ but with $\alpha=0.05$, (c) same as in Fig.~\ref{fig:linear_response} with $s=1$ but with $\alpha=0.05$}
    \label{fig:s1_weak}
\end{figure}

In the main text, we observed deviations between the variational master equation calculations and the tensor network calculations. We attributed those deviations to the strong coupling of the untransformed slow modes of the phonon bath. In this appendix, we show calculations with $\alpha = 0.5 \Delta / (\nu_c \Gamma(s))$ rather than $\alpha = \Delta / (\nu_c \Gamma(s))$ for $s=1$, and observe much better agreement with the tensor network and the variational ME. In Fig.~\ref{fig:s1_weak}(a), we show the population evolution $\expval{\sigma}_z$, which is still captured well. In Fig.~\ref{fig:s1_weak}(b), we show the coherences $\expval{\sigma}_x$, where the deviation between the variational ME and tensor network calculations is now much smaller, albeit still present. The weak coupling is still far off. Finally, in Fig.~\ref{fig:s1_weak}(c), we show the linear absorption and also have a much better agreement.

\section{Linear response for Markovian model \label{app:linear_response}}
In this appendix, we show that for a simple Markovian model, the linear response in Eq.~\eqref{eq:linear_response} vanishes.

We take

\begin{equation}
H_S = \frac{\Delta}{2}\sigma_x ,
\end{equation}

and include a simple Markovian dephasing term,

\begin{equation}
\frac{d\rho}{dt}
=
\mathcal L \rho
=
-i\left[\frac{\Delta}{2}\sigma_x,\rho\right]
+
\gamma_\phi \mathcal D_{\sigma^\dagger\sigma}[\rho],
\end{equation}

where

\begin{equation}
\mathcal D_A[\rho]
=
A\rho A^\dagger
-
\frac{1}{2}\left \{ A^\dagger A,\rho\right \}.
\end{equation}

Writing the general TLS density matrix as

\begin{equation}
\rho(t)
=
\frac{1}{2}
\left[
a_0(t)\mathbb I
+
a_x(t)\sigma_x
+
a_y(t)\sigma_y
+
a_z(t)\sigma_z
\right],
\end{equation}

one obtains

\begin{align}
\dot a_0 &= 0,\\
\dot a_x &= -\frac{\gamma_\phi}{2}a_x,\\
\dot a_y &= -\frac{\gamma_\phi}{2}a_y-\Delta a_z,\\
\dot a_z &= \Delta a_y .
\end{align}

The important observation is that the $\sigma_x$ component is dynamically decoupled from the $\sigma_y,\sigma_z$ sector. Therefore, a perturbation which initially has no $\sigma_x$ component can never acquire one under the Markovian evolution generated by $\mathcal L$.

The linear-response function corresponding to a weak perturbation proportional to $\sigma_x$ is

\begin{equation}
\chi_{xx}(\tau)
=
-i\tr {
\sigma_x e^{\mathcal L\tau}
[\sigma_x,\rho_{\mathrm{ss}}]
},
\end{equation}

To evaluate the initial condition for this auxiliary evolution, we write

\begin{equation}
\rho_{\mathrm{ss}}
=
\frac{1}{2}
\left[
\mathbb I
+
x_{\mathrm{ss}}\sigma_x
+
y_{\mathrm{ss}}\sigma_y
+
z_{\mathrm{ss}}\sigma_z
\right].
\end{equation}

Then

\begin{equation}
[\sigma_x,\rho_\mathrm{ss}]
=
i y_{\mathrm{ss}}\sigma_z
-
i z_{\mathrm{ss}}\sigma_y .
\end{equation}

Thus, the QRT auxiliary operator starts entirely in the $\sigma_y,\sigma_z$ sector and has no $\sigma_x$ component. Since the Markovian Liouvillian above does not transfer $\sigma_y$ or $\sigma_z$ components into $\sigma_x$, it follows that

\begin{equation}
\tr{ \sigma_x e^{\mathcal L\tau}
[\sigma_x,\rho_{\mathrm{ss}}]}
=
0
\end{equation}

for all $\tau$. Consequently,

\begin{equation}
\chi_{xx}(\tau)=0,
\qquad
\chi_{xx}(\omega)=0 .
\end{equation}

This shows that, for the Markovian reference problem, the linear response of $\sigma_x$ vanishes. 

For the full problem including phonons, the non-Markovian effects associated with this coupling, however, mean that the $\sigma_x$ component will couple to the $\sigma_y,\sigma_z$ components, and we thus predict a non-zero linear response, due to the non-Markovianity present. 

\section{Error curves \label{app:error_curves}}

\begin{figure}[!ht]
    \centering
    \includegraphics[width=\linewidth]{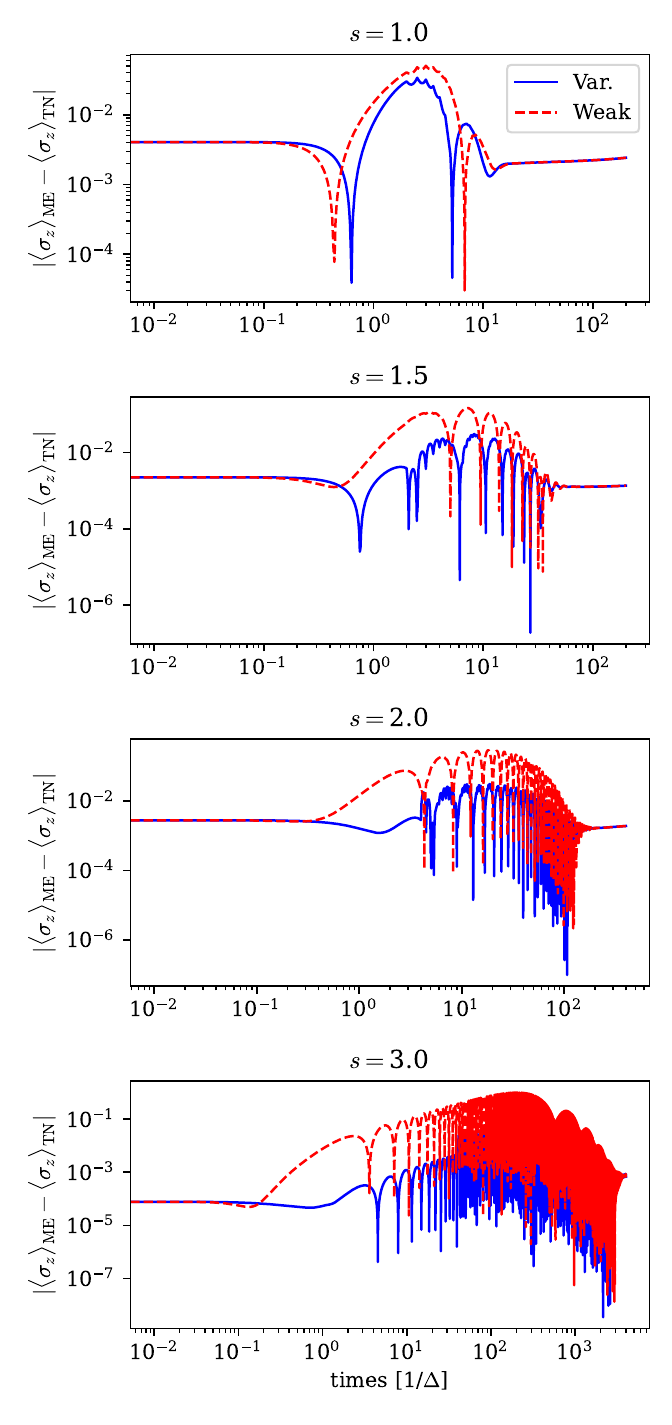}
    \caption{Same as Fig.~\ref{fig:excited_false} but instead the deviation between the master equation approaches, and the tensor network calculations are shown.}
    \label{fig:err_z}
\end{figure}

In this appendix, we provide the results of Figs.~\ref{fig:excited_false}, \ref{fig:ground}, and \ref{fig:linear_response}, but where we instead show the deviation of the master equation results when compared with the tensor network calculations. In Fig.~\ref{fig:err_z}, we show the deviation for the populations, and in general, it is clear that the error of the variational approach is smaller than that of the weak master equation.

\begin{figure}[!ht]
    \centering
    \includegraphics[width=\linewidth]{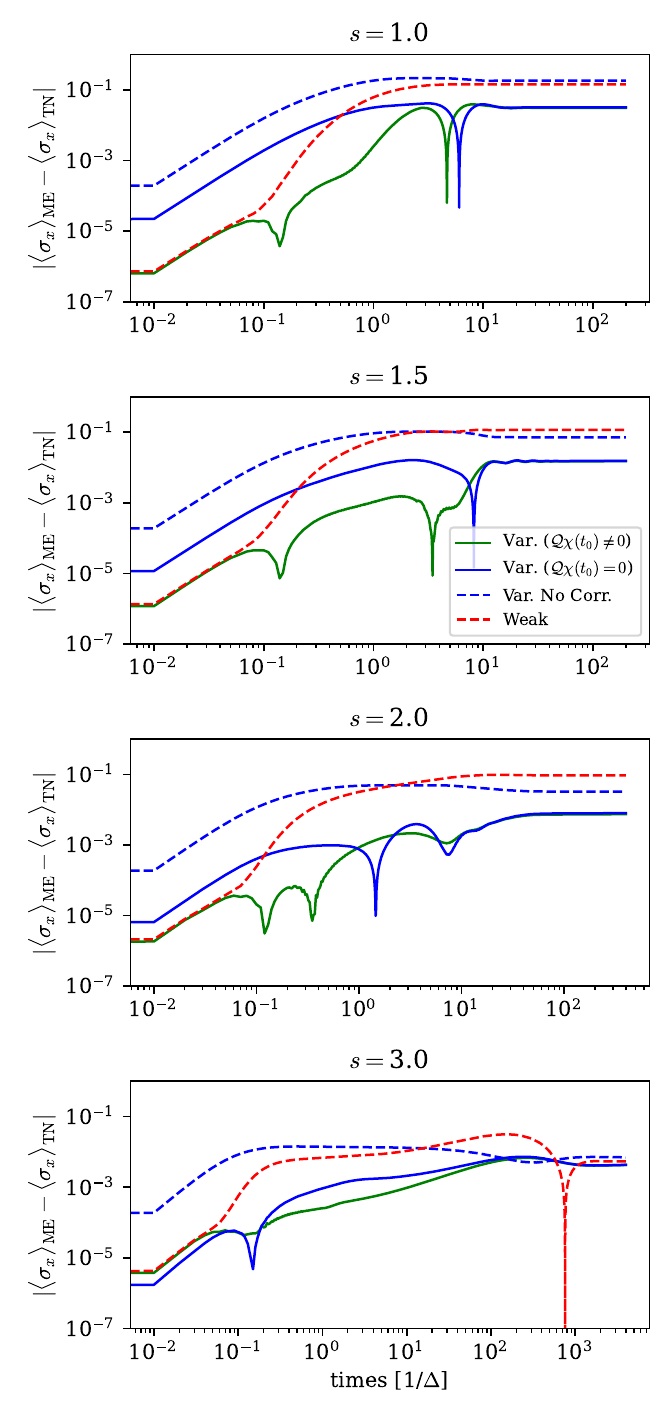}
    \caption{Same as Fig.~\ref{fig:ground} but instead the deviation between the master equation approaches, and the tensor network calculations are shown. For the weak master equation and the Variational master equation, including the inhomogeneous terms ($\mathcal Q \chi(t_0) \neq 0$) the error is computed against the non-relaxed tensor network calculations. For the variational master equation approach without inhomogeneous terms ($\mathcal Q \chi(t_0) = 0$) with and without correction terms, the error is computed against the relaxed tensor network calculations. }
    \label{fig:err_x}
\end{figure}

In Fig.~\ref{fig:err_x}, we show the error of the coherence $\expval{\sigma_x}$, where again it is clear that the variational approach including corrections leads to much smaller errors, especially in the steady state. Note that we here compare the weak master equation and the Variational master equation, including the inhomogeneous terms ($\mathcal Q \chi(t_0) \neq 0$) against the non-relaxed tensor network calculations. While for the variational master equation approach without inhomogeneous terms ($\mathcal Q \chi(t_0) = 0$) with and without correction terms, the error is computed against the relaxed tensor network calculations.

\begin{figure}[!ht]
    \centering
    \includegraphics[width=\linewidth]{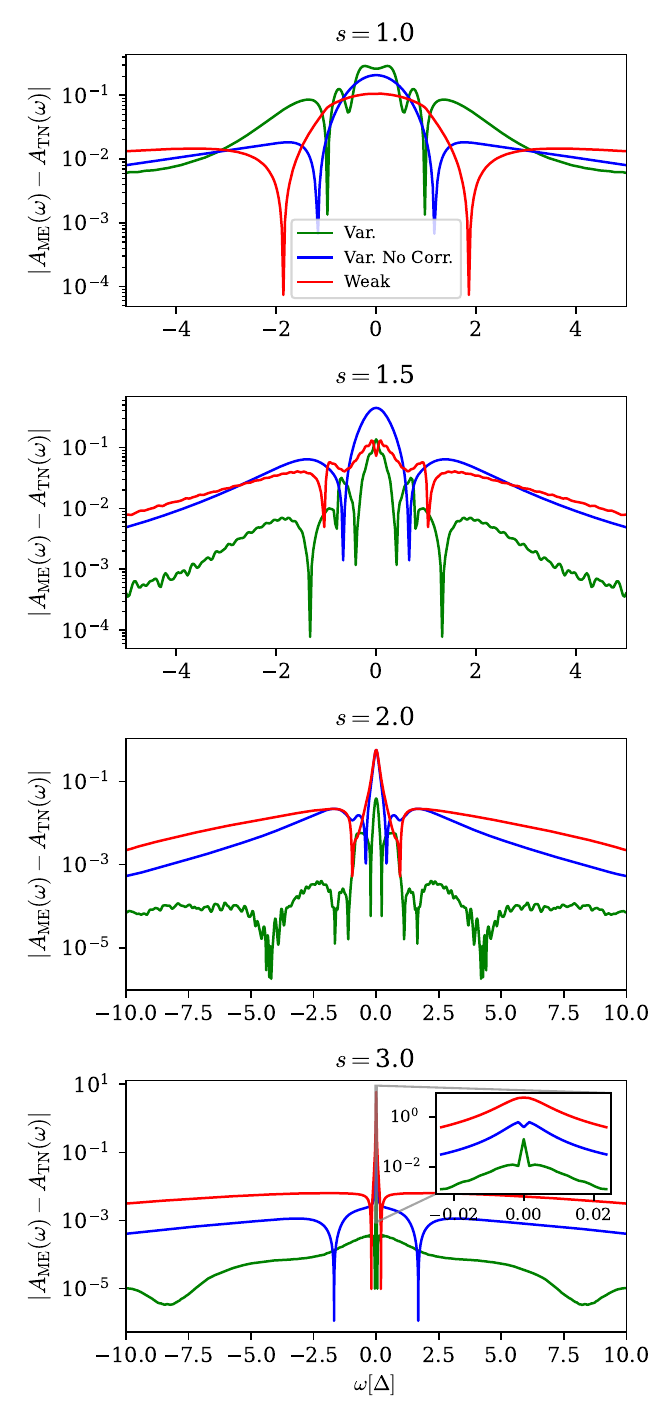}
    \caption{Same as Fig.~\ref{fig:linear_response} but showing the deviation from the tensor network calculations. }
    \label{fig:err_linear}
\end{figure}

In Fig.~\ref{fig:err_linear}, we show the error of the coherence linear response, and here it is clear that generally the error is smaller if the corrections to the regression theorem are included, both in the weak and variational approaches.

%

\end{document}